\documentclass[a4paper,11pt]{article}
\pdfoutput=1 

\usepackage{jheppub} 

\usepackage[T1]{fontenc} 
\usepackage{graphicx}

\newcommand{\alinea}{\hspace*{\parindent}}

\def\krto{ {\,\,\lower .8ex\hbox {$\longrightarrow \atop k \rightarrow 0$}\,\,}}

\def\alinea{\hspace{\parindent}}
\def\bea{\begin{eqnarray} }
\def\beq{\begin{eqnarray} }

\def\eea{\end{eqnarray}}
\def\eeq{\end{eqnarray}}

\def\eq#1{Eq.~(\ref{#1})}

\newcommand{\VG}{\langle G^2 \rangle}
\def\eq#1{Eq.~(\ref{#1})}
\def\eqs2#1#2{Eqs.~(\ref{#1},\ref{#2})}
\def\Bp#1#2#3{\widetilde{B}_{#1}^{#2}({\bf #3})}
\def\P#1#2#3{\left(\delta^{#1#2} - \frac{#3_{#1}#3_{#2}}{#3^2}\right)}

\title{\boldmath Instanton liquid properties from lattice QCD}

\author[a]{A.~Athenodorou} 
\author[b]{, Ph.~Boucaud} 
\author[c,h,1]{, F.~De Soto \note{Corresponding author}}
\author[d,h]{, J.~Rodr\'{\i}guez-Quintero}
\author[e,f,g ]{and S.~Zafeiropoulos.}

\affiliation[a]{Computation-based Science and Technology Research Center, Cyprus Institute, 20 Kavafi Str., Nicosia 2121, Cyprus}

\affiliation[b]{Laboratoire de Physique Th\'eorique (UMR8627), CNRS,
Univ. Paris-Sud, Universit\'e Paris-Saclay, 91405 Orsay, France}

\affiliation[c]{Dpto. Sistemas F\'{\i}sicos, Qu\'{\i}micos y Naturales, 
Univ. Pablo de Olavide, 41013 Sevilla; Spain}

\affiliation[d]{Dpto. Ciencias Integradas, Fac. Ciencias Experimentales; 
Universidad de Huelva, 21071 Huelva; Spain.}

\affiliation[e]{Institute for Theoretical Physics, Heidelberg University,
Philosophenweg 12, 69120 Heidelberg, Germany}
\affiliation[f]{Department of Physics, The College of William \& Mary, Williamsburg, VA 23187, USA}
\affiliation[g]{Thomas Jefferson National Accelerator Facility, Newport News, VA 23606, USA}

\affiliation[h]{CAFPE, Universidad de Granada, E-18071 Granada, Spain}

\emailAdd{a.athenodorou@cyi.ac.cy}
\emailAdd{philippe.boucaud@th.u-psud.fr}
\emailAdd{fcsotbor@upo.es}
\emailAdd{jose.rodriguez@dfaie.uhu.es}
\emailAdd{zafeiropoulos@thphys.uni-heidelberg.de}

\abstract{
We examined the instanton contribution to the QCD configurations generated from lattice QCD for $N_F=0$, $N_F=2+1$ and $N_F=2+1+1$ dynamical quark flavors from two different and complementary approaches. First via the use of Gradient flow, we computed instanton liquid properties using an algorithm to localize instantons in the gauge field configurations and studied their evolution with flow time. Then, the analysis of the running at low momenta of gluon Green's functions serves as an independent confirmation of the instanton density which can also be derived without the use of the Gradient flow.}

\begin{document}
\maketitle
\flushbottom


\section{Introduction}
\alinea

Instantons and other semi-classical solutions of the QCD Lagrangian are believed to play an essential role in the low energy dynamics of QCD, where the crucial phenomena of confinement and chiral symmetry breaking take place~\cite{RevModPhys.70.323}. 
Instantons are by definition topologically non-trivial solutions
of the classical field equations in Euclidean space with finite
action. These solutions play an interesting role from Quantum Mechanics, where they describe 
tunneling processes, in Minkowski spacetime, from one vacuum to another vacuum all the way to Yang-Mills theories where they describe tunnelling processes between different vacua which are labeled by a different value of the topological charge (winding number). Additionally, they play a crucial role in the explanation of the mechanism of spontaneous breaking of chiral symmetry~\cite{Diakonov:1995ea} but their relation to confinement in four dimensional theories is much less clear
despite the initial success in three-dimensional gauge theories~\cite{Polyakov:1976fu}.
Apart from tunnelling paths in Minkowski spacetime, instantons are intimately related to many interesting phenomena such as the $U(1)$ problem, the strong $CP$ problem, they have interesting counterparts in the electroweak sector (sphalerons) that can lead to violation of the baryon and lepton number conservation and could possibly describe rare processes of baryon decay. Also, in supersymmetric gauge theories the exact $\beta$-function can be computed by instanton methods. For more details we refer the interesting reader to~\cite{Vandoren:2008xg}.

QCD topology is a fully non-perturbative topic, and naturally the best way to study it systematically is in ab-initio lattice simulations. Strictly speaking, topology is not mathematically well defined on the lattice. One can employ different definitions of the topological charge either gluonic (which are based on some discretization of the topological charge density after some smoothing procedure) or fermionic (based on the spectrum of the Dirac operator). However, for small values of the lattice spacing were the gauge fields are smooth enough, it is still meaningful to study the topology of gauge fields and there are intensive explorations in  lattice simulations~\cite{PhysRevD.89.105005,PhysRevD.92.125014,Cichy:2014yca,Alexandrou:2017hqw}. While not demonstrated theoretically, it is expected and also confirmed by numerical simulations that all definitions of the topological charge agree in the continuum limit (of course can be affected by different sizes of cutoff effects)~\cite{Alexandrou:2017hqw}.

The main reason that a smoothing procedure is needed when one employs a gluonc definition of the topological charge is that both the topological charge and the instanton contribution to the lattice gauge fields are ``hidden" by the presence of short-range (UV) fluctuations and most studies reveal the presence of instantons only after the application of such a filtering technique. Cooling~\cite{Teper:1985rb}, different smearing methods and, more recently, Gradient flow are efficient techniques that remove short-range fluctuations in the gauge fields that have been widely exploited for the study of QCD topology~\cite{PhysRevD.92.125014}. 

After UV fluctuations have been removed, different methods have been used to recognize instantons in the remaining gauge fields~\cite{PhysRevD.52.4691,PhysRevD.78.054506,PhysRevD.88.034501} and measure their density and size distributions.
The application of a filtering technique may nevertheless introduce biases in the characteristics of the underlying semiclassical configuration. For example, instanton/anti-instanton pair annihilation would lead to a reduction of the instanton density that would therefore depend on the amount of UV filtering applied. The fact that the instanton size may be modified by the filtering is also known, at least with the use of cooling, in such a way that depending on the gauge action used, instantons shrink or grow with cooling. A third phenomenon, the disappearance of small instantons of size comparable to the lattice spacing may introduce uncontrolled and worse effects because it would affect the value of the topological charge. 

A different approach to the determination of the instanton nature of QCD vacuum was presented in \cite{JHEP04Boucaud}, where the IR running of some gluon Green's functions was asserted to be related to some properties of the instanton ensemble. Much effort has been devoted by the non-perturbative QCD community to the understanding of the deep IR running of the correlation functions among the fundamental fields of the theory. Quite remarkably, the combination of results from lattice, Dyson-Schwinger equations and some other continuum approaches has led to the firm conclusion that the gluon propagator acquires a dynamical mass in the deep IR while the ghost propagator remains massless~\cite{Cucchieri:2007md,Cucchieri:2007rg,Frasca:2007uz,Boucaud:2008ji,Boucaud:2008ky,Aguilar:2008xm,Dudal:2008sp,Bogolubsky:2009dc,Oliveira:2010xc,Tissier:2010ts,Aguilar:2012rz,Oliveira:2012eh,Ayala:2012pb,Pelaez:2014mxa,Gao:2017uox}. An appealing possibility to describe the large distance (low momenta) correlations among gluon fields is the use of an instanton liquid model with the advantage that it can be applied both before and after the removal of UV fluctuations. 

Our proposal in this paper is to apply both the direct recognition of instantons and the analysis of the running of gluon Green's functions, using the comparison between both methods in order to quantify the systematic uncertainties present in each method. We will furthermore analyze the instanton properties before applying the Gradient flow from i) an extrapolation of the results to zero flow time and ii) the analysis of the IR running without Gradient flow. Overall we expect to obtain a precise image of the instanton description of the QCD vacuum, and its evolution with Gradient flow.
Using this combination of methods we will explore the dependence of the instanton liquid parameters on the number of dynamical fermions of the theory, by exploiting quenched ensembles ($N_F=0$) and unquenched ensembles (with $N_F=2+1$ and $N_F=2+1+1$ dynamical flavors).

The action density and topological charge density obtained after removing UV fluctuations show the presence of large lumps that are thought to be the realization of the IR structure of the initial gauge fields. One possible description of those structures are instantons. Section \ref{sec:classic} will be devoted to a brief description of the Ans\"atze commonly used for the instantonic description of the QCD vacuum. A fit of the topological charge density obtained from the lattice to the instanton prediction has been used to quantify not only the topological charge but the instanton density as well (see section \ref{sec:locating}).

Although instantons are believed to be the basis of long-distance physics of the QCD vacuum, few studies have searched instanton traces without the application of any filtering technique. A few years ago, the IR running of the gluon Green's functions was proposed as a way to study the instantonic content of lattice gauge field configurations without filtering short-range fluctuations~\cite{ATHENODOROU2016354,PhysRevD.70.114503,JHEP03Boucaud}. 

The aim of the paper is twofold. The fist purpose of the paper is therefore to combine the methods based in the analysis of the IR running of gluon Green's functions and the localization of instantons by fitting the lumps after application of the Gradient flow. The second motivation is to describe the parameters of the instanton liquid model before the application of any filtering technique.
This article starts with the definition of the multi-instanton background which is introduced in Section 2, along with all the Landau gauge Green's functions formulae of the instanton liquid model. In Section 3 the local instanton recognition method is introduced along with all the basics of the Gradient flow. In Section 4 we present a detailed analysis of the IR running of 
the MOM coupling constant, with and without the effects of the Gradient flow. In the same Section the fate of instantons, under the Gradient flow, is studied in a simple toy model. Section 5 summarizes our conclusions.


\section{Multi-instanton background}
\label{sec:classic}

Our main purpose is to make the semiclassical multi-instanton background resulting from the non-trivial QCD vacuum manifest, basically by computing its contribution to the gluon gauge field, assuming it to be dominant in the deep infrared domain, and identifying then this contribution from results obtained from lattice simulations. In order to do so, we can proceed in two different manners: (i) a direct recognition of local structures 
in configuration space either for the gauge action or for the topological charge density, due to the instanton brackground, and (ii) by the scrutiny of the low-momenta gluon correlations expressed by the appropriate Green's functions. Let us then start by introducing a practical description of the gauge field within the classical multi-instanton background.

\subsection{The gauge field within a multi-instanton ensemble}

We base our description of the QCD vacuum in a reliable approximation of a multi-instanton solution for the $SU(3)$ gauge action, built on the ground of the appropriate Ansatz for the minimization of the resulting action per particle. The following trial function, 
\beq\label{r-ansatz}
g_0 B_\mu^a({\bf x}) \ = \ \frac {\displaystyle 2 \sum_{i=I,A} R_{(i)}^{a\alpha} \overline{\eta}^\alpha_{\mu\nu} 
\frac {y_i^\nu}{y_i^2} \ \rho_i^2 \frac{f(|y_i|)}{y_i^2} } 
{\displaystyle 1 + \sum_{i=I,A} \rho_i^2 \frac{f(|y_i|)}{y_i^2}} \ ,
\eeq
has been proposed in ref.~\cite{Shuryak:1987iz} for the gauge-field classical solution from an ensemble of instantons, $B_\mu^a$; where $y_i=({\bf x}-z^i)$ and $\overline{\eta}^\alpha_{\mu\nu}$ is the 't~Hooft tensor, that should be replaced by $\eta^\alpha_{\mu\nu}$ when summing over anti-instantons as $i=A$. $R_{(i)}^{a\alpha}$ represents the color rotations embedding the canonical $SU(2)$ instanton solution in the $SU(3)$ gauge group ({\it i.e., $\alpha=1,2,3$} and $a=1,2, \dots 8$). $f(x)$ is a shape function, to be obtained by the minimization of the action, that should obey $f(0)=1$ in order not to spoil the field topology at the instanton centers and which additionally provides sufficient cut-off at large distances guaranteeing convergence of the sum. The length parameter $\rho$ is known as the instanton radius or size and, of course, remains transparent for a classical solution of the field equations. It can be only fixed, phenomenologically, after a successful comparison of the instanton-based prediction of a given observable with its empirical result, or related to the lattice spacing when describing lattice results. In terms of this length scale, regimes of large and small distances can be asymptotically defined from \eq{r-ansatz} and, as discussed in \cite{ATHENODOROU2016354}, in both cases, the gauge field can be effectively described by the following independent-pseudoparticle Ansatz~\cite{Diakonov:1983hh}
\begin{equation}
g_0\, B_{\mu}^{a}({\bf x})= 2 \sum_i R^{ a \alpha}_{(i)}\,
\overline{\eta}^\alpha_{\mu \nu} \ \frac {y_i^\nu}{y_i^2} \
\phi_{\rho_i}\left(\frac{|y_i|}{\rho_i}\right) \ , \label{eq:sumansazt}
\end{equation} 
which appears to be a linear superposition of pseudo-instantons at different positions, and where the profile function $\phi$ behaves as
\bea\label{eq:match}
\phi_\rho(z) = \left\{ 
\begin{array}{lr} 
\displaystyle \frac{f(\rho z)}{f(\rho z)+z^2} \simeq \frac 1 {1+z^2} & z \ll 1
 \\
\displaystyle \frac{f(\rho z)}{z^2} & z \gg 1
\end{array} \right. \ ,
\eea
where $f(x)$ is the shape function defined in \eq{r-ansatz}. As will be seen below, a linear superposition makes possible a simple and closed result for the gluon two- and three-point correlation functions, reliable both in the large as in the small-distance domains.  
The profile function $\phi$ appears to match the behaviors from both domains, expressed by the shape function, and breaks explicitly the scale independence required by the (small-distance) limit of an isolated instanton, provided by the BPST solution~\cite{BELAVIN197585}
\beq
\phi_{\rm BPST}(z)=\frac{1}{1+z^2} \, .
\eeq
Thus, assuming a BPST profile near any instanton center, one can therein approximate the topological charge density by
\beq
Q_{\rm BPST}(x)= \pm \frac{6}{\pi^2\rho^4} \left(\frac{\rho^2}{(x-x_0)^2+\rho^2}\right)^4 \, ,
\label{eq:bpst}
\eeq
with ``$+$'' sign for instantons and ``$-$'' sign for anti-instantons. 

On the other hand, while the one-instanton contributions dominate at small distances (near any instanton center) over the non-linear effects from other instantons in the background, the latter are dominant for large distances and their average define the drop of the shape function which, according to \cite{Diakonov:1983hh}, can be approximated as being also independent of the instanton size. 

It is worth emphasizing that, even when using a profile function $\phi(z)$ that takes into account instanton interactions, the Ans\"atze given by Eqs.~\eqref{r-ansatz} and \eqref{eq:sumansazt} are only approximations to the solutions for the classical equations of motion.\footnote{Beyond the well-known BPST solution for an isolated instanton, only a few more particular cases admitting a known closed expression are known, as the n-instanton configuration (see \cite{atiyah} for a detailed treatment).} Later below, aiming at a careful examination of the classical background underlying the gauge fields produced by lattice QCD simulations, we will apply a prescription based on the local recognition of the peaks induced in the topological charge density by the pseudo-instantons of \eq{eq:sumansazt}. Such a goal requires that both Ans\"atze provide a good description of the gauge field locally around each pseudo-instanton center, such that the topological charge density results fairly approximated by Eq.~\eqref{eq:bpst}. Furthermore, a non-local prescription for the scrutiny of the classical background in the gauge field, based on the analysis of two- and three-point gluon Green's functions, has been recently investigated in \cite{ATHENODOROU2016354} and will be extensively used in the following. For this latter method to work, \eq{r-ansatz} needs to be a good approximative representation for the classical gauge field within the multi-instanton background, and the deviations in \eq{eq:sumansazt} with respect to \eq{r-ansatz} are to be neglected from the gluon correlations in momentum space, at least in the low-momentum regime. Its main advantage is however that, as will be later explained, it can be directly applied to the lattice gauge fields without any filtering to deprive the fields from quantum fluctuations; being thus a cheaper and less distorting prescription for the examination of the quasi-classical background.

\subsection{Green's functions in Landau gauge}

Let us assume the independent pseudo-particle Ansatz in (\ref{eq:sumansazt}) for the gauge field within the multi-instanton ensemble, {\it i.e.} an instanton liquid model (ILM) for the gauge field. Then, if we assume that instanton positions and color orientations are uncorrelated, the gluon propagator dressing function reads~\cite{Boucaud:2002fx,JHEP03Boucaud,Boucaud:2003xi} 
\beq\label{eq:G2}
G^{(2)}(k^2) \, = \, \frac{1}{24} \delta_{ab} \left(\delta^{\mu\nu}-\frac{k^\mu k^\nu}{k^2}\right) \langle \Bp{\mu}{a}{k} \Bp{\nu}{b}{-k} \rangle 
\, = \, \frac{n}{8g_0^2} \langle \rho^6 I^2(k\rho) \rangle \ , 
\eeq
in Landau gauge (always in Euclidean space), with $\sqrt{V} \widetilde{B}_\mu^a({\bf k}) = \int \mathrm{d}^4{\bf k} e^{i {\bf k} \cdot {\bf x}} B_\mu^a({\bf x})$, $V$ being formally the volume of spacetime and where the usual average over the gauge group space, $\langle\cdots \rangle$, is recast by the ILM as the average over all the possible configurations of pseudo-instantons within the statistical ensemble defining the classical background of the QCD vacuum. All the dependence on the profile function $\phi(z)$ is captured by 
\beq
I(s) = \frac{8\pi^2}{s} \int_0^\infty dz\ z J_2(s z) \phi(z) \ ,
\eeq
where $J_2$ is a Bessel function of the first kind. Analogously, albeit requiring more algebra, the scalar form factor for the tree-level component of the symmetric Landau-gauge three-gluon Green's function,   
\beq
G^{(3)}(k^2) &=& \langle \Bp{\mu}{a}{p} \Bp{\nu}{b}{q} \Bp{\rho}{c}{r} \rangle \, f_{abc} \, \nonumber \\
&\times & \left[ \Gamma^{(0)}_{\mu' \nu' \rho'}\left({\bf p},{\bf q},{\bf r}\right)  \P{\mu'}{\mu}{p} \P{\nu'}{\nu}{q} \P{\rho'}{\rho}{r} 
\right. \nonumber \\  \label{eq:G3}
&& \left. + \, \frac 1 2 \, \frac{(p-q)^\rho (q-r)^\mu (r-p)^\mu}{p^2} \right] \  \\
\mathrm{where} & & 
\Gamma^{(0)}_{\mu \nu \rho}\left({\bf p},{\bf q},{\bf r}\right)  \, = \, \delta_{\mu \nu} \left(p_\mu - q_\nu\right) + 
\delta_{\nu \rho} \left( q_\nu - r_\rho \right)  + \delta_{\rho \mu} \left( r_\rho  - p_\mu\right) \ ,
\eeq
reads~\cite{Boucaud:2002fx,JHEP03Boucaud,Boucaud:2003xi} 
\beq
G^{(3)}(k^2) &=& \frac{n}{48 k g_0^3} \langle \rho^9 I^3(k\rho) \rangle \ ,
\eeq
the kinematical configuration of momenta defined by $p^2=q^2=r^2=k^2$. 

A quite remarkable feature of the strong coupling defined in the so-called symmetric MOM scheme~\cite{Boucaud:1998bq}, computed from the form factors \eqref{eq:G2} and \eqref{eq:G3}, 
\beq\label{eq:amom}
\alpha_{\rm MOM}(k) \, = \, \frac{k^6}{4\pi} \frac{\left(G^{(3)}(k^2)\right)^2}{\left(G^{(2)}(k^2)\right)^3}
\, = \,  \frac{k^4}{18\pi n} \, \frac{\langle \rho^9 I(k\rho)^3\rangle^2}{\langle \rho^6 I(k\rho)^2\rangle^3} \ ,
\eeq
happens to be its independence on the instanton profile in both the limits of small and large momenta~\cite{JHEP04Boucaud,ATHENODOROU2016354}, as a result of
\beq
 \frac{\langle \rho^9 I(k\rho)^3\rangle^2}{\langle \rho^6 I(k\rho)^2\rangle^3}
\, = \, \left\{\begin{array}{l}
1 + \mathcal{O}\left(\displaystyle \frac{\delta\rho^2}{k^2\bar{\rho}^4} \right) \\
\displaystyle 1 + 48 \frac{\delta\rho^2}{\bar{\rho}^2} +  \mathcal{O}\left(k^2\delta\rho^2,\frac{\delta\rho^4}{\bar{\rho}^4}\right) 
\end{array} \right. \ ,
\label{eq:k4}
\eeq
where $\bar\rho=\sqrt{\langle \rho^2\rangle}$ expresses the mean instanton size and $\delta\rho^2=\langle (\rho - \bar\rho)^2\rangle$ stands for the mean square width of the instanton size distribution. Then, at large and small momenta, the contribution from the classical instanton background to the running of the coupling defined by \eq{eq:amom} remains insensitive to both the instanton profile function and to the statistical distribution of the multi-instanton ensemble represented by the independent pseudo-particle Ansatz in (\ref{eq:sumansazt}). Indeed, it will only keep track of the instanton density, as immediately results from plugging \eqref{eq:k4} into \eq{eq:amom}, one thus obtaining that the coupling approximately behaves according to a $k^4$-power law wherein the classical background appears to be 
dominant. This was clearly shown in Ref.~\cite{JHEP04Boucaud} to happen at low momenta, where the ultraviolet (UV) quantum fluctuations have little effect on the gluon correlations, by a careful examination of the coupling computed from lattice QCD. The same also happens, after filtering the short-range UV fluctuations out from the gauge fields, at large momenta~\cite{JHEP03Boucaud,Boucaud:2004zr,ATHENODOROU2016354}.  In particular in \cite{ATHENODOROU2016354}, we applied the so-called Gradient flow as filtering technique and thus harvested a solid confirmation of the picture emerging from \eqs2{eq:amom}{eq:k4}. This is especially so, because the effect of the size-distribution width yields a different factor in front of the power for low and large momenta (see \eq{eq:k4} above), and this is precisely what the analysis made in  \cite{ATHENODOROU2016354} appeared to show. 
We will supplement here this non-local technique, in some representative cases, with a local one based on the shape recognition of peaks in the topological charge density which, albeit more expensive from the point of view of computing time, will be of great help for pinpointing the instanton sizes and obtaining a complete picture of the instantonic description.

\section{Instanton local recognition from lattice QCD} 
\label{sec:lattice}

As explained in the previous subsection, one of the main objectives of this paper is to examine the lattice gauge fields in configuration space, in order to reveal their underlying multi-instanton content by applying the local recognition recipe that will be introduced and discussed below. A consistent comparison of the properties so obtained for the quasi-classical multi-instanton ensemble with those extracted with the Green's function method will be finally presented.  

\subsection{The topological charge density}

As already advanced, the examination of the classical background underlying the lattice gauge fields in lattice configurations is based on the recognition of the shape of the peaks, locally around their centers, induced in the topological charge density by the pseudo-particles employed to build the multi-instanton Ansantz for the gluon gauge field, as given by \eq{eq:sumansazt}. The first step needs thus to be the computation of the topological charge density, which is obtained, as well as the gauge action, as follows.

Let us call $\Pi_{\mu\nu}({\bf x})$ the Hermitean traceless part of the plaquette starting at site ${\bf x}$ in the  $\mu-\nu$ plane, 
\begin{equation}\label{eq:plaquette}
\Pi_{\mu\nu}({\bf x}) = \frac{\square_{\mu\nu}({\bf x})-\square^\dagger_{\mu\nu}({\bf x})}{2} - \frac{1}{3}{\rm Tr}\frac{\square_{\mu\nu}({\bf x})-\square^\dagger_{\mu\nu}({\bf x})}{2}\ ,
\end{equation}
with $\square_{\mu\nu}({\bf x})$ the average of the four plaquettes in the $\mu-\nu$ plane. Then, the action density can be obtained as 
\begin{equation}
S({\bf x}) = \frac{1}{8\pi^2} \sum_{\mu>\nu} {\rm Tr}  \left[ \Pi_{\mu\nu}({\bf x})^2 \right] \ ,
\end{equation}
where the $8\pi^2$ normalization factor comes from the action of a single instanton, such that $\int \mathrm{d}^4{\bf x} S({\bf x})=1$ either for the instanton or  anti-instanton cases. While, for the topological charge density one is left with
\begin{equation}\label{eq:topological-charge}
Q({\bf x}) = \frac{1}{2^5\pi^2}  \sum_{\mu\nu\rho\sigma} {\rm Tr}  
\left[ \epsilon_{\mu\nu\rho\sigma}\Pi_{\mu\nu}({\bf x})\Pi_{\rho\sigma}({\bf x})\right] \ ,
\end{equation}
which, again, keeps the appropriate normalization factor to result in $\int d^4x Q({\bf x})=+1 (-1)$ for instantons (anti-instantons).
\eq{eq:topological-charge} is the so-called clover definition of the topological charge density, which is expressed by a combination of the plaquettes given in \eq{eq:plaquette}. More elaborated definitions, which include rectangular Wilson loops of size $1\times 2$ in order to reduce discretization errors, can be found in literature (see, {\it e.g.} \cite{Alexandrou:2017hqw}). Notwithstanding this, our use of \eq{eq:topological-charge}, supplemented with the prescription and acceptance criteria that will be described below, as well as with the comparison with the Green's function results, will be enough to achieve our purpose of revealing the quasi-classical multi-instanton background underlying the gauge fields and making its properties manifest.  

However, the quantum UV fluctuations are obviously present in any ensemble of lattice gauge fields obtained by Monte Carlo techniques and implementing the QCD action, therefore hiding the quasi-classical background. In computing the gluon correlations, the UV fluctuations have been shown to be negligible in the low-momentum regime, basically dominated by the long-distance correlations~\cite{JHEP03Boucaud,JHEP04Boucaud,ATHENODOROU2016354}. On the contrary, gauge fields are locally dominated by short-range contributions, which, if they are not previously suppressed, would spoil the topological charge density defined by \eqref{eq:topological-charge}. 
In the following subsection, we will briefly explain the use of the Gradient flow to remove such short-range contributions, as it has been also done in~\cite{ATHENODOROU2016354}, in that case to unveil the quasi-classical $k^4$-power law also at large-momenta.

\subsection{Gradient flow}
\label{sec:WF}

The Gradient flow can be conceived as a smoothing procedure which diminishes the short-distance fluctuations which, within the context of a quantum field theory, correspond to UV quantum fluctuations. Therefore, depriving the gauge fields from them, potentially,
implies to isolate the underlying non-trivial classical solutions which minimize the gauge action. Despite of sharing the capacity of eliminating the short-distance fluctuations with other filtering techniques such as cooling or smearing, the Gradient flow has a theoretical ground with attractive features such as the simple renormalization of the flown fields~\cite{Luscher2011}. 

In continuum language, the Gradient flow field $B_\mu(\tau,{\bf x})$ of $SU(N)$ gauge field results from the solution of the following first order differential equation
\beq
\frac{\partial B_\mu}{\partial \tau} = D_\mu G_{\mu\nu}\ ;
\eeq
that introduces the flow time $\tau$ and
\beq\label{eq:flown}
G_{\mu\nu} &=& \partial_\mu B_\nu - \partial_\nu B_\mu + \left[ B_\mu, B_\nu \right]  \ , \nonumber \\
D_\mu & = & \partial_\mu + \left[ B_\mu, \cdot \right] \ ,
\eeq
the field strength tensor for the flown fields and covariant derivative, respectively. As initial condition for the differential equation, one chooses $B_\mu(0,{\bf x})=A_\mu({\bf x})$, so that the fundamental gauge field $A_\mu({\bf x})$ evolves from $\tau=0$ and, flown after any $\tau$, reads 
\beq
B_\mu(\tau,{\bf x})=\int \mathrm{d}^4{\bf y} \; \frac{\displaystyle e^{-\frac{({\bf x}-{\bf y})^2}{4\tau}}}{(4\pi\tau)^2} A_\mu({\bf x}) \ ,
\eeq
as a formal expansion that, at tree-level, makes apparent an effective suppression of short-distance fluctuations up to a distance of $\sqrt{8\tau}$. Its lattice formulation, the $SU(N)$ matrices are flown by obeying a discretized counterpart of the first-order differential equation \eqref{eq:flown}, 
\beq
V_\mu(\tau,{\bf x}) \ = \ - g_0^2 \left[ \partial_{x,\mu} S(V(\tau)\right] \; V_\mu(\tau,{\bf x}) \ , 
\eeq
with the initial condition $V_\mu(0,{\bf x})=U_\mu({\bf x})$, and where $S$ is a discretization of the gauge action (herein, we will mainly use the Wilson gauge action for this, and will only make use of the Iwasaki gauge action to perform some tests, as discussed below), $g_0$ is the bare coupling and $\partial_{x,\mu}$ are the link derivatives that can be found defined in~\cite{Luscher2010}. Despite its important advantages as it is possessing a well defined continuum limit or the existence, uniqueness and smoothness of its solutions, the effect of Gradient flow over the gauge fields is in the practical level equivalent to that of cooling, at least for the purpose of dealing with the topological charge~\cite{PhysRevD.89.105005,PhysRevD.92.125014,Alexandrou:2017hqw}.
A remarkable property of the Gradient flow is however the steadiness of any exact solution of $D_\mu G_{\mu\nu}=0$ under the flow, thereupon implying that quantum UV fluctuations can be filtered out without distorting exact solutions such as an isolated instanton. In practice, this property of the Gradient flow does not prevent it from modifying the parameters of the multi-instanton ensemble \eqref{eq:sumansazt}, representing the quasi-classical background underlying the lattice gauge fields. And the latter is true for two different reasons: first because the whole argument about steadiness corresponds to the continuum formulation of both instantons and flow, and second because the multi-instanton representations are only approximated solutions to the classical equations of motion, the isolated instantons never being an adequate description of the QCD vacumm, where a rather dense distribution is expected~\cite{PhysRevD.88.034501}.

For comparative purposes, the flow time should be expressed in units of $t_0$, defined by $\sqrt{8t_0}=0.3 {\rm fm}$ following~\cite{Luscher2010}. This allows to settle physical units for the Gradient flow times, given that $t_0=a^2\tau_0=0.01125 {\rm fm}^2$ and $t=\tau \; t_0 /\tau_0$.

\subsection{Locating instantons}
\label{sec:locating}

Let us start by describing the lattice simulations set-ups for the gauge field configurations both quenched ($N_F=0$) and unquenched (with $N_F=2+1$ and $N_F=2+1+1$ dynamical flavors) that will be either exploited in this section by applying the instanton local recognition, for some representative cases, or subsequently {\it via} the scrutiny of the low-momenta running of the coupling defined by \eq{eq:amom}. 

For the quenched data, we consider two simulations exploited in \citep{ATHENODOROU2016354} and \cite{PhysRevD.95.114503} with lattice spacings and volumes appropriate for having both enough lattice sites in the vicinity of an instanton peak and enough data for momenta within the IR window, after Fourier transform. 
In the case of unquenched simulations, we will use two sets of lattice data, the first includes $N_F=2+1$ field configurations produced by the RBC/UKQCD collaboration using domain wall fermions very close to the physical pion mass ($139$ MeV), more details on the lattice set-up can be found in \cite{PhysRevD.93.074505}; the second set corresponds to a $N_F=2+1+1$ simulation from the ETM collaboration with a pion mass of around $300$ MeV. This last simulation has been carried out using Wilson twisted mass fermions at maximal twist (details of the lattice setup can be found in \cite{Baron2010}). Details such as the gauge action, the values of the lattice spacing, the spacetime volumes or the total number of exploited configurations for each set-up appear gathered in Tab.~\ref{tab:lattice}. 

As will be seen below, we will take the configurations for the tree-level Symanzik gauge action at $\beta$=4.2 (see Tab.~\ref{tab:lattice}) as a representative 
case and apply the Gradient flow for $\tau=2,3,4,5,8,10$ and $15$. 
Only for these fields, we will apply the instanton local recognition, while the Green's function method will be applied for all of the lattice ensembles displayed in Tab.~\ref{tab:lattice}. It should be particularly noticed that, by the combination of different gauge actions in the quenched case, we intend to guarantee that the discretization effects are under control or, at least, that the observed behavior appears irrespectively of the action considered. Finally, having access to $N_F=0$, $N_F=2+1$ and $N_F=2+1+1$ dynamical flavor configurations we will be in a position to check the dependence on the instanton liquid parameters with the number of fermionic flavors in the sea.

\begin{table}[th]
\begin{center}
\begin{tabular}{c c c c c c c c}
\hline
\hline
$N_F$ & gauge action & fermion action & $\beta$ & $a$ & $V$  & confs. & ref. \\
\hline
\hline
0  & tlS  &  --  & 4.2     & $0.141~{\rm fm}$        & $(4.5 ~{\rm fm})^4$        & 420   & \cite{ATHENODOROU2016354}  \\ 
  &  W   &      & 5.8     & $0.140~{\rm fm}$        & $(6.72~{\rm fm})^4$       & 960   &  \\
\hline
2+1 & lw & DWF   & 2.25     & $0.0835~{\rm fm}$        & $5.34^3 \times 10.7~{\rm fm}^4$       & 330  &  \cite{PhysRevD.93.074505}  \\
       &       &      & 2.13     & $0.1139~{\rm fm}$        & $5.48^3 \times 11.0~{\rm fm}^4$       & 350  &   \\
\hline
2+1+1 & Iw & TMF & 1.95     & $0.083~{\rm fm}$        & $4.0^3\times 7.9~{\rm fm}^4$       &   200 & \cite{Baron2010}\\
\hline
\hline
\end{tabular}
\end{center}
\caption{Set-ups for the different ensemble of lattice data that have been exploited for this paper. Concerning the codes for the actions, tlS stands for tree-level Symanzik, W for Wilson, Iw for Iwasaki, DWF for Domain-Wall fermions and TMF for twisted-mass fermions.}
\label{tab:lattice}
\end{table}

Now, after applying the Gradient flow, the short distance fluctuations that are present in both the action and topological charge densities are suppressed and smooth lumps, presumably induced by the pseudo-particles in \eq{eq:sumansazt}, appear unveiled. We will then explicitly check for those lumps in the topological charge density, locally around their centers,  their shape similarity to the classical instanton solution in Eq.~(\ref{eq:bpst}). In the rest of this section we will discuss the algorithm we have used for identifying instantons in the flown field configurations. The first step is to localize the extrema of the topological charge density that will be candidates for topological charges. Then, an estimate for their sizes has to be provided. The aim of the last step will be to eliminate candidates that are not topological charge structures but remain UV fluctuations.

We will define local maxima (minima) of the topological charge density as those sites $x_0$ where $Q(x_0)$ is larger (smaller) than the closest 8 neighbors. We are aware that defining the extrema with respect to the 8 sites that are at $\pm a$ away in one direction may produce double counting when a single instanton is so much distorted by quantum fluctuations that another extremum takes place close the real one in the surrounding hypercube. Had we used a more strict definition of an extremum, such as $Q(x)$ being larger than the 80 sites of the surrounding hypercube (see discussion in \cite{Smith:1998wt}), we would have avoided the double counting but would have been also left to disregard the distorted real instanton. Therefore, we have preferred to use a more modest definition of an extremum and apply a further filter to eliminate false candidates at the last step.  

\begin{figure}[h!]
\begin{center}
\begin{tabular}{c} 
\begin{tabular}{cc}
\includegraphics[width=0.52\textwidth]{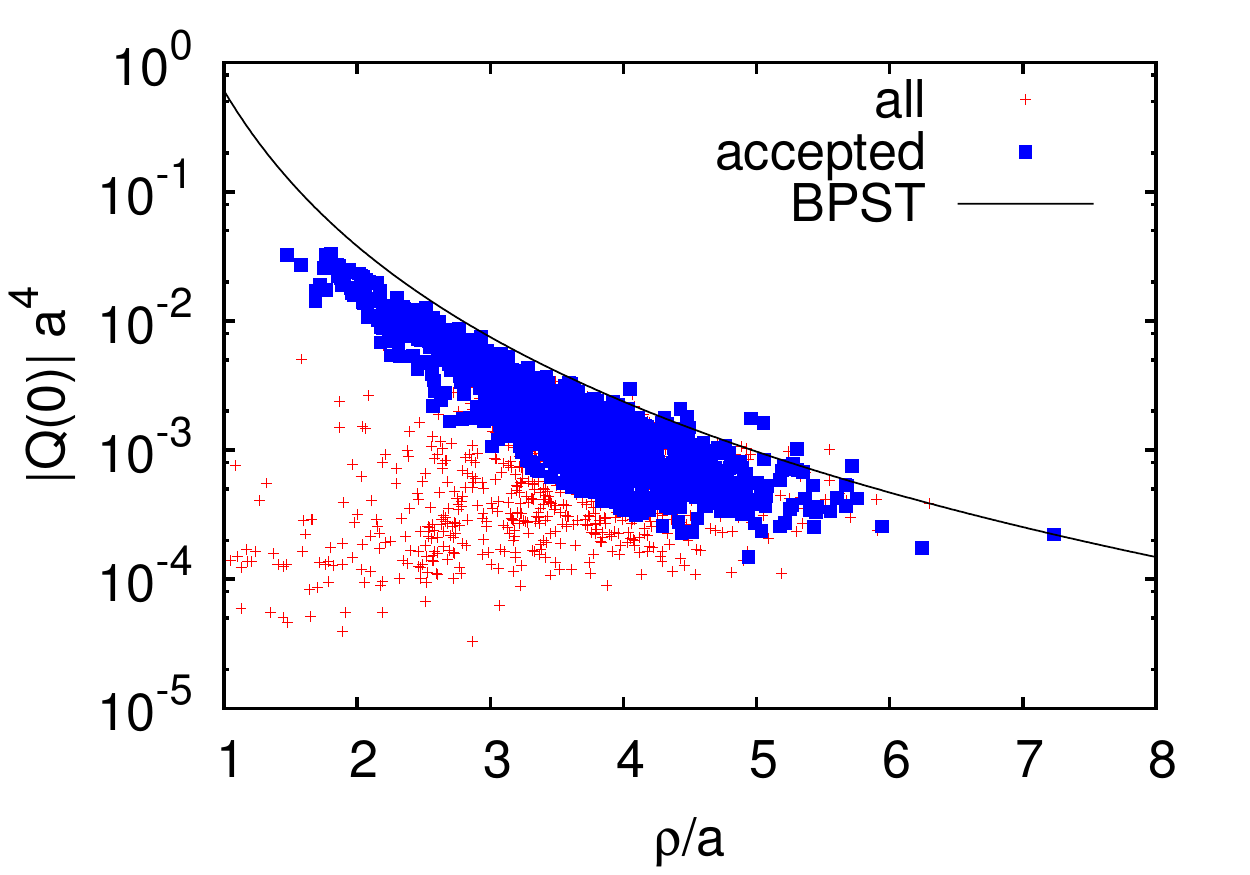} &
\includegraphics[width=0.52\textwidth]{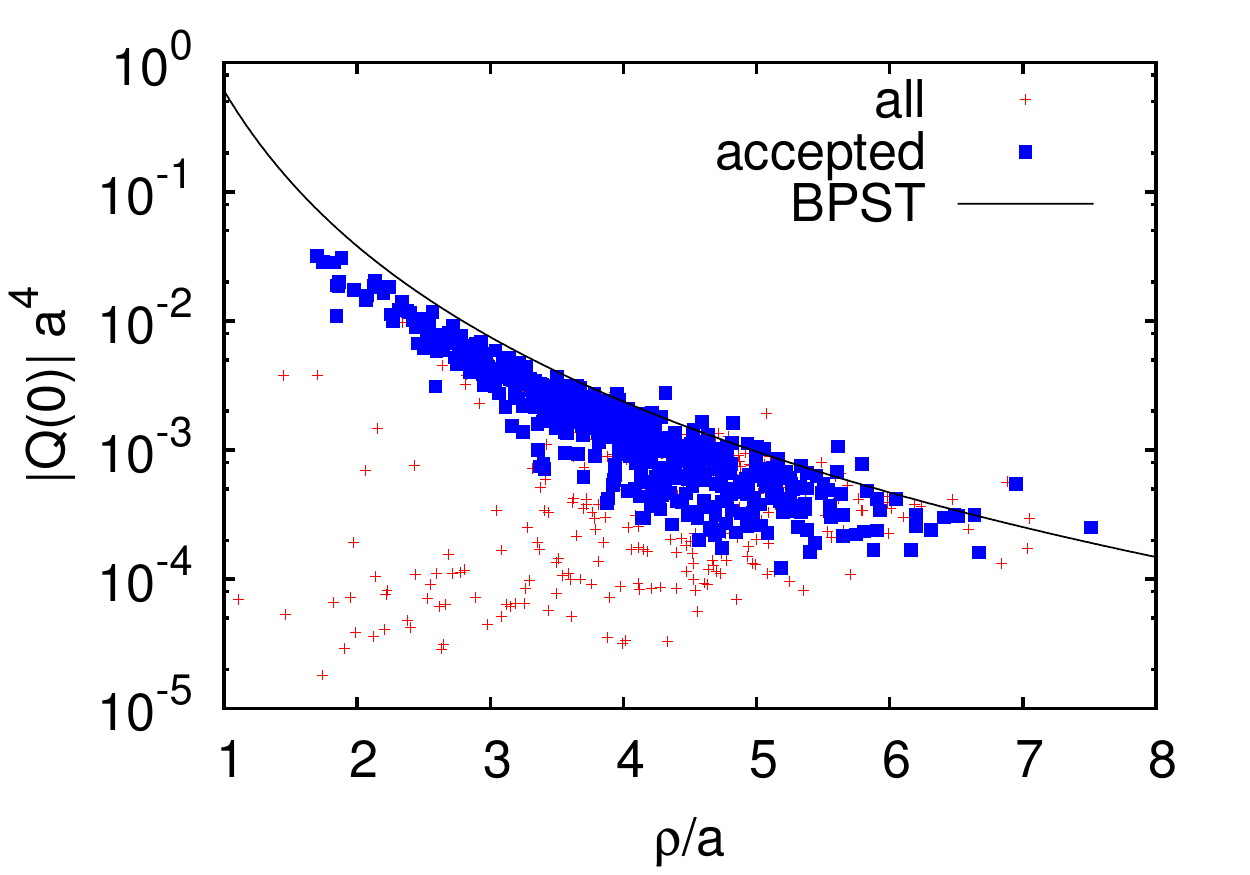} 
\end{tabular} 
\\
\includegraphics[width=0.52\textwidth]{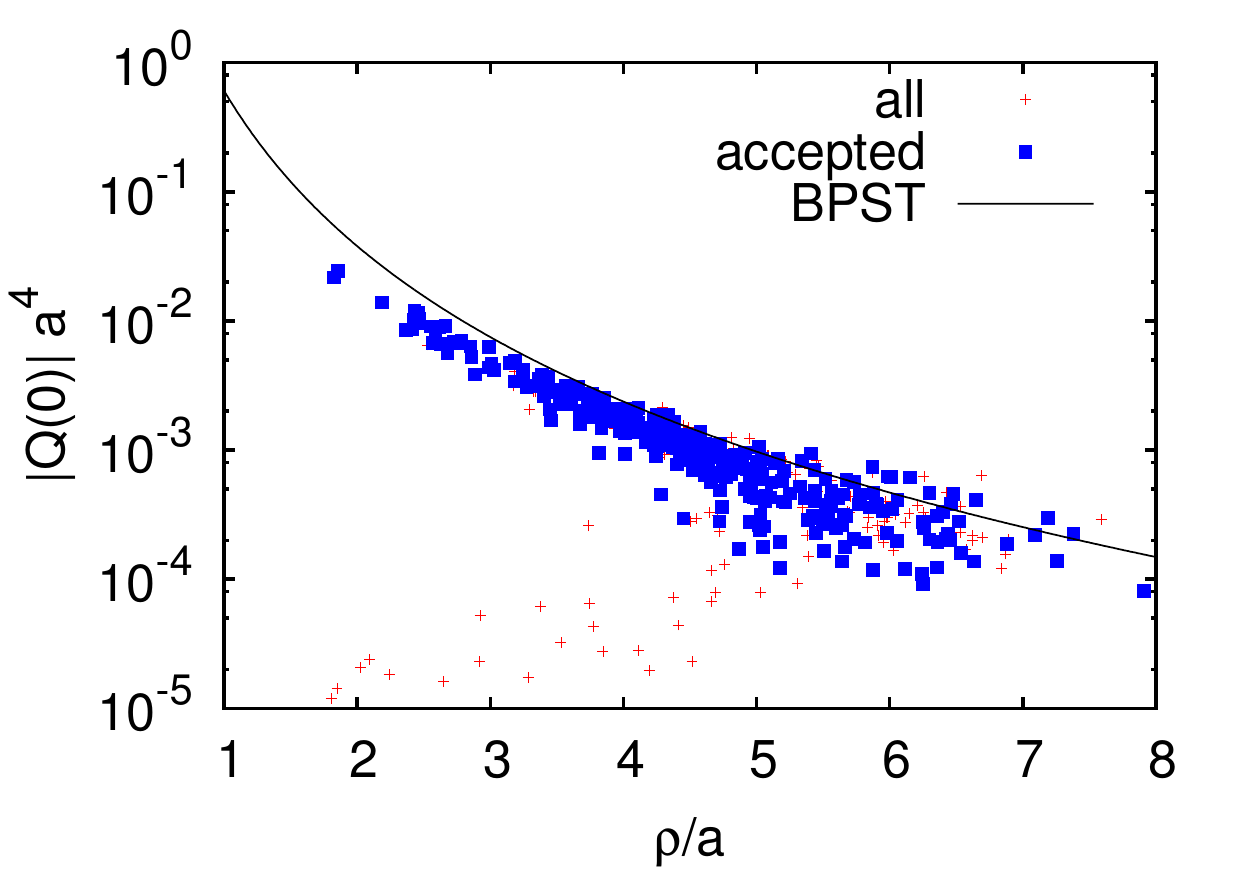}
\end{tabular}
\end{center}
\caption{(Color online) Absolute value of the topological charge density at the extrema (in lattice units) vs. instanton size $\rho$ determined from the fit described in the text for a particular gauge field configuration of $\beta=4.2$, $V=32^4$ after $\tau$=4 (leftmost upper panel), 8 (rightmost upper) and 15 (lower) Gradient flow times. Those extrema accepted as instantons are represented by blue squares, while the rejected ones are represented by red crosses. The full line represents the relation \eq{eq:bpst}.}
\label{fig:q_vs_rho}
\end{figure}

After localizing these instanton candidates, the next step will be to obtain their instanton size $\rho$. To this aim, we consider the ratio of the topological charge at the closest 8 neighbors to the value at the center that, according to  Eq.(\ref{eq:bpst}) should be
\beq
\frac{Q({\bf x})}{Q({\bf x_0})} = \left(\frac{\rho^2}{({\bf x}-{\bf x_0})^2+\rho^2}\right)^4
\eeq
and fitted the lattice data to this expression using $\rho$ as a free parameter. According to Eq.~(\ref{eq:bpst}), the value of the topological charge at the center of a single instanton is related to the instanton size by $Q(x_0)=\frac{6}{\pi^2\rho^4}$. To check this dependence, the peak value of the topological charge density has been plotted versus the fitted size in Fig.~\ref{fig:q_vs_rho} for three typical quenched configurations after $\tau$=4, 8 and 15  Gradient flow times. In this plot there is a point for each local extremum of the topological charge density, while the continuum line corresponds to Eq.~(\ref{eq:bpst}). As can be seen, most of the extrema lie near the continuum line  although a non-negligible number of outliers have to be filtered out as described below.  Furthermore, the larger is the flow time, the more closely concentrated around the continuum line tend the extrema to be. 

Then, as a first filtering prescription, we proceed first by checking that the topological charge and action densities for all the candidates behave as \eq{eq:sumansazt} for short distances, and reject those not satisfying
\beq\label{eq:Qoverrho4}
\frac{\sqrt[4]{\frac{6}{\pi^2 Q({\bf x_0})}}}{\rho} \in (1-\epsilon_R,1+\epsilon_R),
\eeq
and
\beq
\frac{|\sum_{{\bf x}={\bf x_0}\pm a {\bf u}} Q(x)|}{\sum_{{\bf x}={\bf x_0}\pm a {\bf u}} S({\bf x})} \in (1-\epsilon_Q,1+\epsilon_Q),
\eeq
where $\epsilon_R$ and  $\epsilon_Q$ are parameters that we fix in our algorithm. We have checked that the total number of instantons found is rather stable when varying them, and thus fixed $\epsilon_R=0.5$ and $\epsilon_Q$=0.3  for the results shown below. Once we ensure that the functional form of the topological charge lump behaves locally as an isolated BPST instanton, and that it is selfdual, we are left with the above raised issue of preventing from the double counting of two extrema corresponding to a single but distorted instanton. To do so, we have finally included a filter for eliminating close pairs that works as follows, when an extremum is identified as an instanton of size $\rho_i$ at site ${\bf x_i}$, no other candidate is accepted if the site of the extremum lies within the hypersphere defined by ${\bf x_i} + \epsilon \rho_i {\bf u}$, with ${\bf u}$ a unitary vector in physical units (and, in all the cases, the hypersphere radius is fixed to 2 in lattice units when the instanton size is smaller). The parameter $\epsilon$ is varied between $0.7$ and $1$, the difference between the densities obtained for those two values being used to estimate the systematic uncertainty associated to our localization method. This filter serves to exclude the  possibility of finding one instanton in the core of another one~\footnote{In other words, we have assumed that, for a bi-instanton solution of the classical field equations, the two extrema cannot appear located at a distance smaller than their individual radius. And so we he have checked for some particular cases, when closed analytical solutions exist.}. The filtering of close pairs eliminates a number of candidates much larger than the functional form of $Q({\bf x})$, but many of them correspond to the outlier extrema of Fig.~\ref{fig:q_vs_rho} that anyhow passed the criterion \eqref{eq:Qoverrho4}, namely those candidates for which instanton size $\rho_i$ and $Q({\bf x_i})$ are not well consistent with \eqref{eq:bpst}.

\begin{figure}[h!]
\begin{center}
\begin{tabular}{c}
\begin{tabular}{c c}
\includegraphics[width=0.5\textwidth]{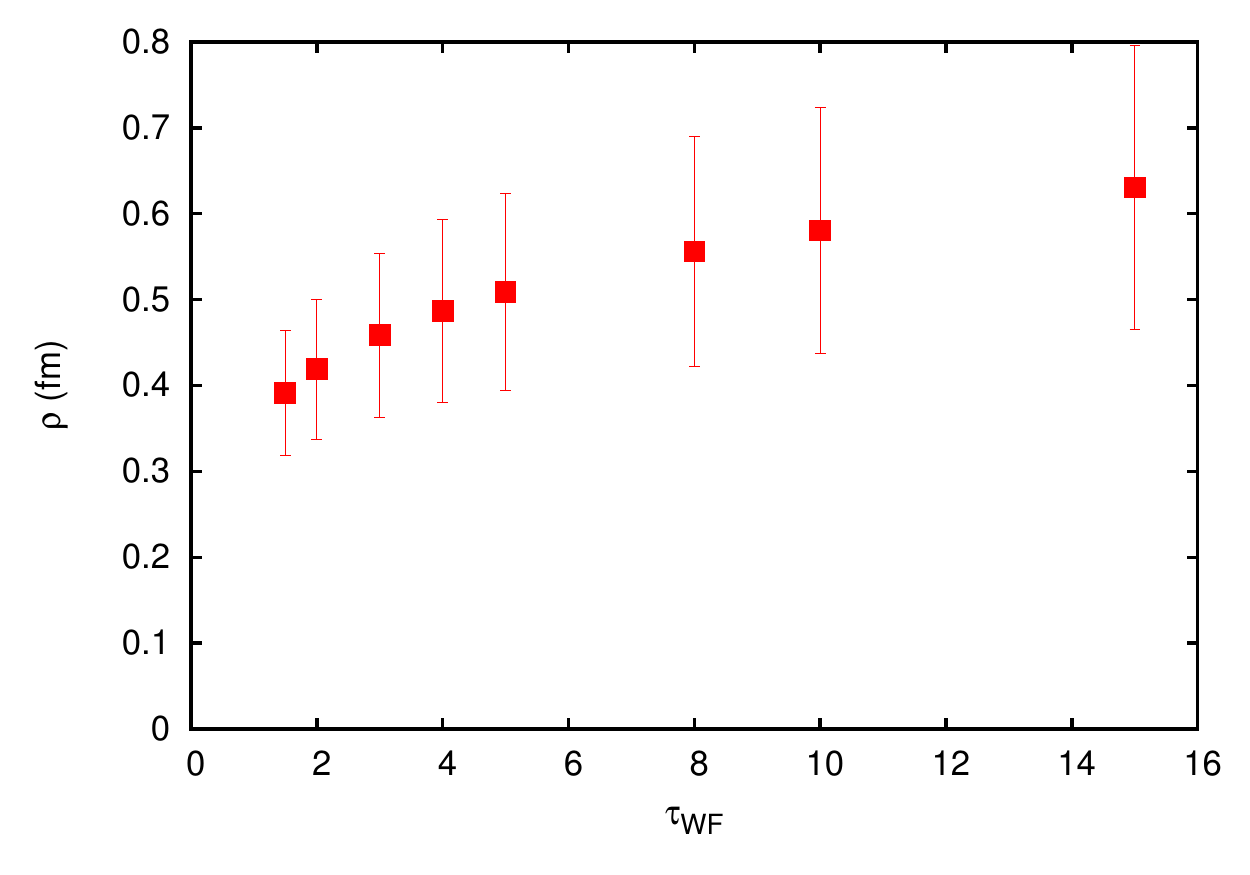} &
\includegraphics[width=0.5\textwidth]{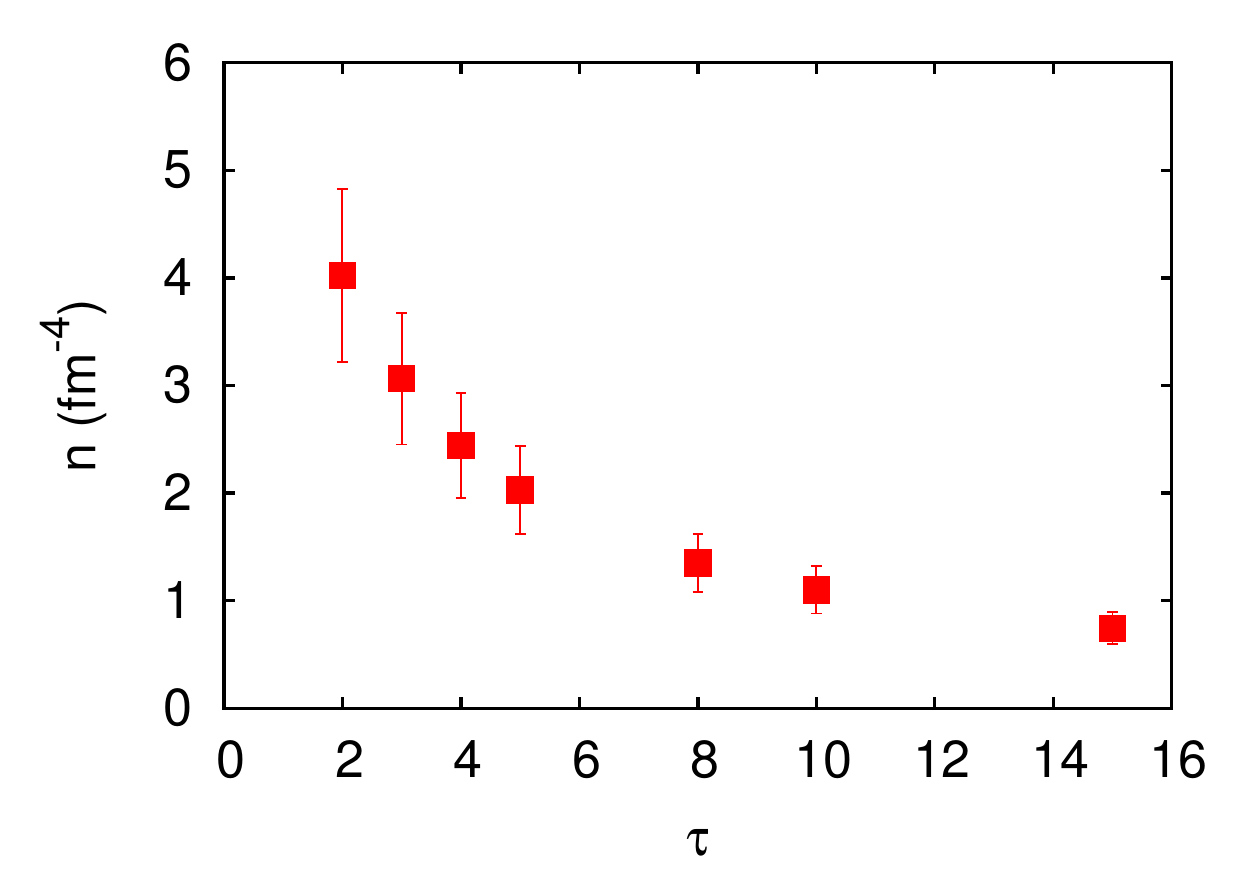} 
\end{tabular} \\
\includegraphics[width=0.5\textwidth]{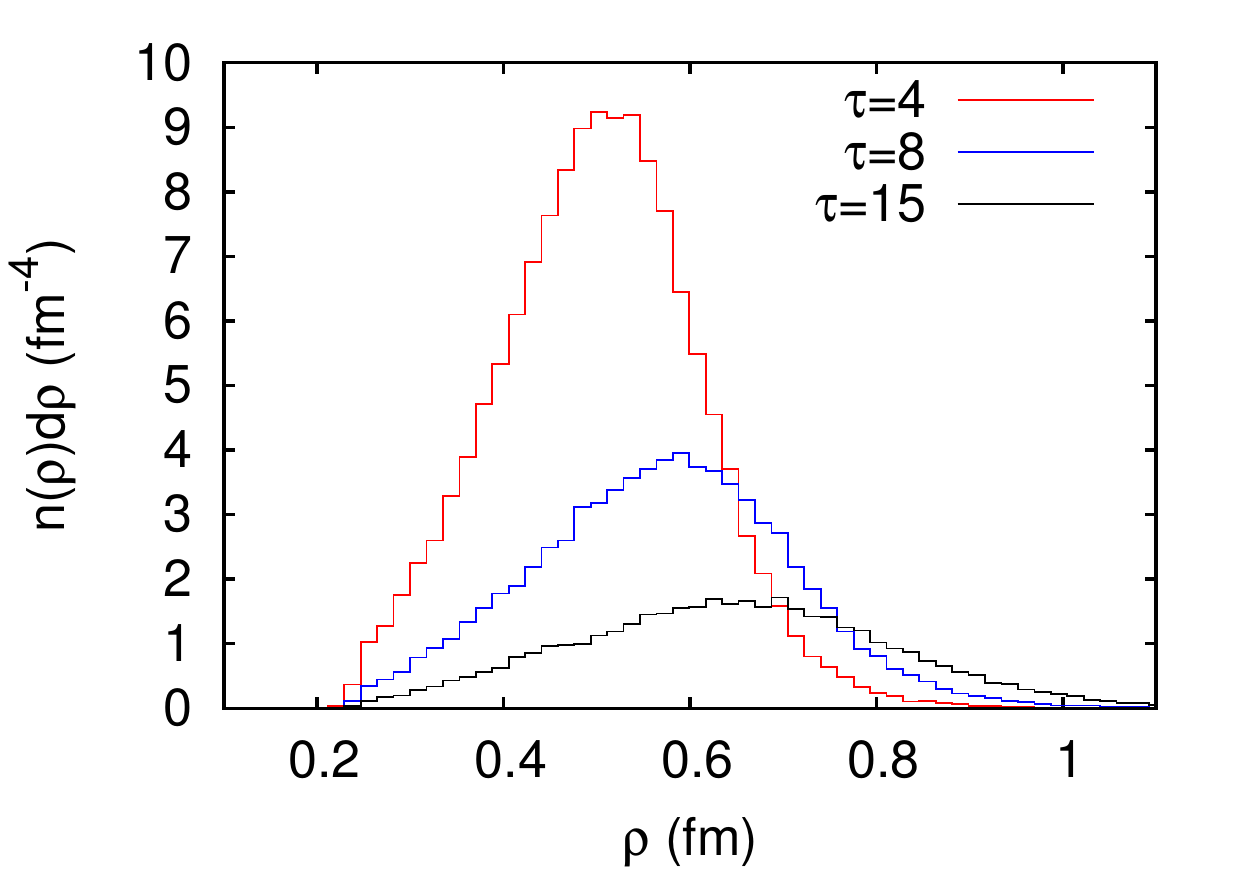} 
\end{tabular}
\end{center}
\caption{(Color online) Evolution of the average instanton size (leftmost upper), density (rightmost upper) and the histogram of sizes (lower) with the Gradient flow time for 100 quenched gauge field configurations corresponding to $\beta=4.2$ (see Tab.~\ref{tab:size_density_Wilson}). The error bars for the instanton size show the width of the distribution rather than the standard deviation of the mean. For the density, the error bars incorporate an estimate of the systematic uncertainty associated to the distance filter discussed in the text.}
\label{fig:size_density_Wilson}
\end{figure}

Both parameters of the instanton ensemble, the average instanton radius and density, which are describing the semiclassical background underlying the gauge field, do depend on the Gradient flow time and, in general, evolve with any parameter controlling the suppression of the quantum fluctuations {\it via} a given smoothing prescription minimizing the action. Although this is well known~\cite{GARCIAPEREZ1994535,NEGELE199992,HART2001280,PhysRevD.88.034501,PhysRevD.89.105005}, 
few authors have studied this evolution in detail, quoting just the values of the instanton density or size distribution after a fixed flow time, cooling or smearing. In our case, after applying the Gradient flow and our instanton locating algorithm, the emerging picture is that of a very dense ensemble of instantons with a size of around 1/3 fm, which is more diluted as Gradient flow eliminates instantons and anti-instantons, while the remaining ones become larger with the flow. This is very apparent from the results displayed in Fig.~\ref{fig:size_density_Wilson} and also in Tab.~\ref{tab:size_density_Wilson}.

\begin{table}[h]
\begin{center}
\begin{tabular}{c c c c c c c}
\hline
\hline
$\tau$ & $t/t_0$ & $n\  {\rm (fm^{-4})}$ & $\bar{\rho}\ {\rm (fm)}$ \\
\hline
\hline
 2  & 3.42 & $4.01(62)$ & $0.419(8)$ \\
 3  & 5.13 & $3.06(43)$ & $0.458(10)$ \\
 4  & 6.84 & $2.44(34)$ & $0.486(11)$ \\
 5  & 8.55 & $2.03(28)$ & $0.509(12)$ \\
 8  & 13.7 & $1.35(18)$ & $0.555(13)$ \\
 10 & 17.1 & $1.10(13)$ & $0.580(14)$ \\
 15 & 25.6 & $0.74(08)$ & $0.639(17)$ \\
\hline
\hline
\end{tabular}
\end{center}
\caption{Instanton density and size for different Gradient flow times in lattice units (first row) and in physical units in terms of $t_0$=0.01125 fm$^2$, obtained from the localization technique. The errors quoted for the instanton density include an estimate of the systematic uncertainty associated to the short distance filter (see text) while the average size errors are purely statistical.}
\label{tab:size_density_Wilson}
\end{table}

\begin{figure}[h!]
\begin{center}
\begin{tabular}{c}
\begin{tabular}{cc}
\includegraphics[width=0.5\textwidth]{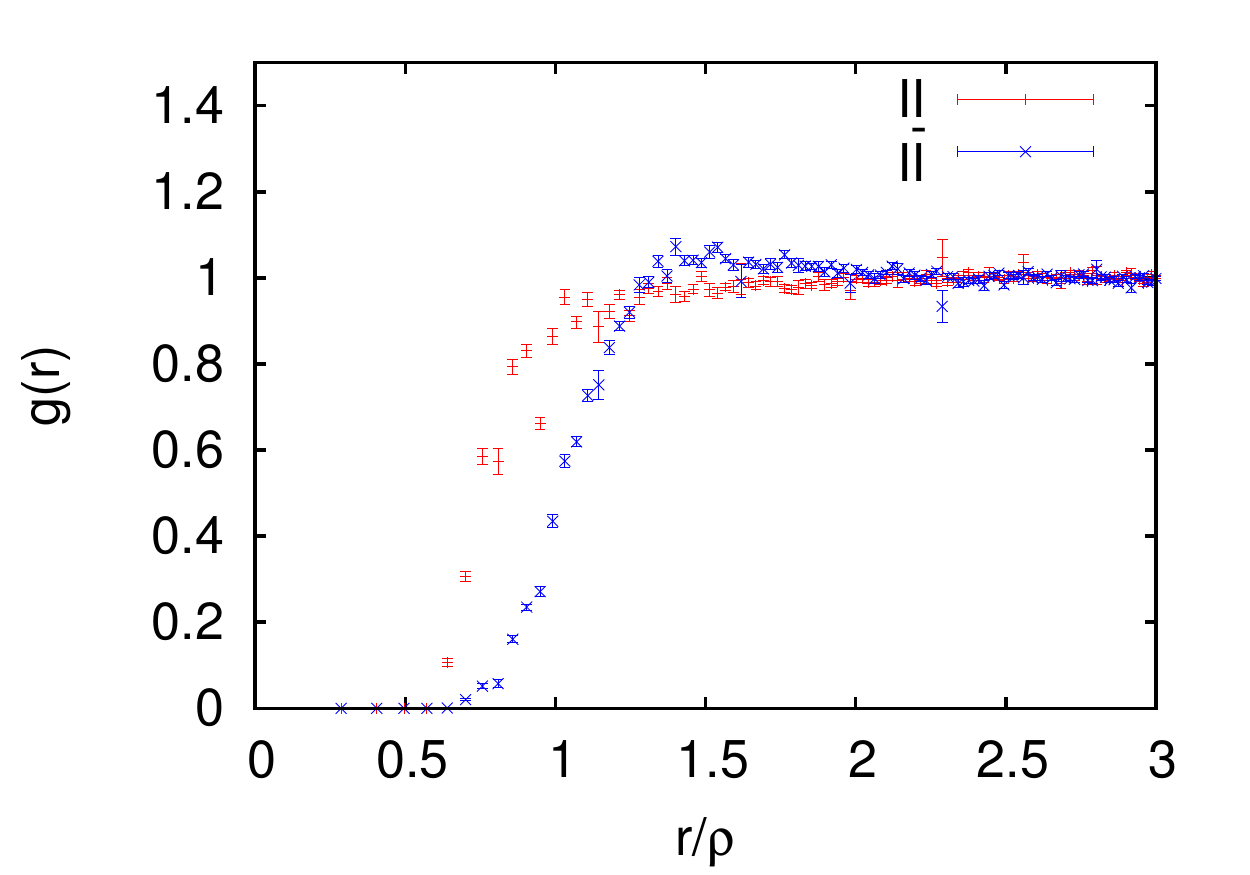} &
\includegraphics[width=0.5\textwidth]{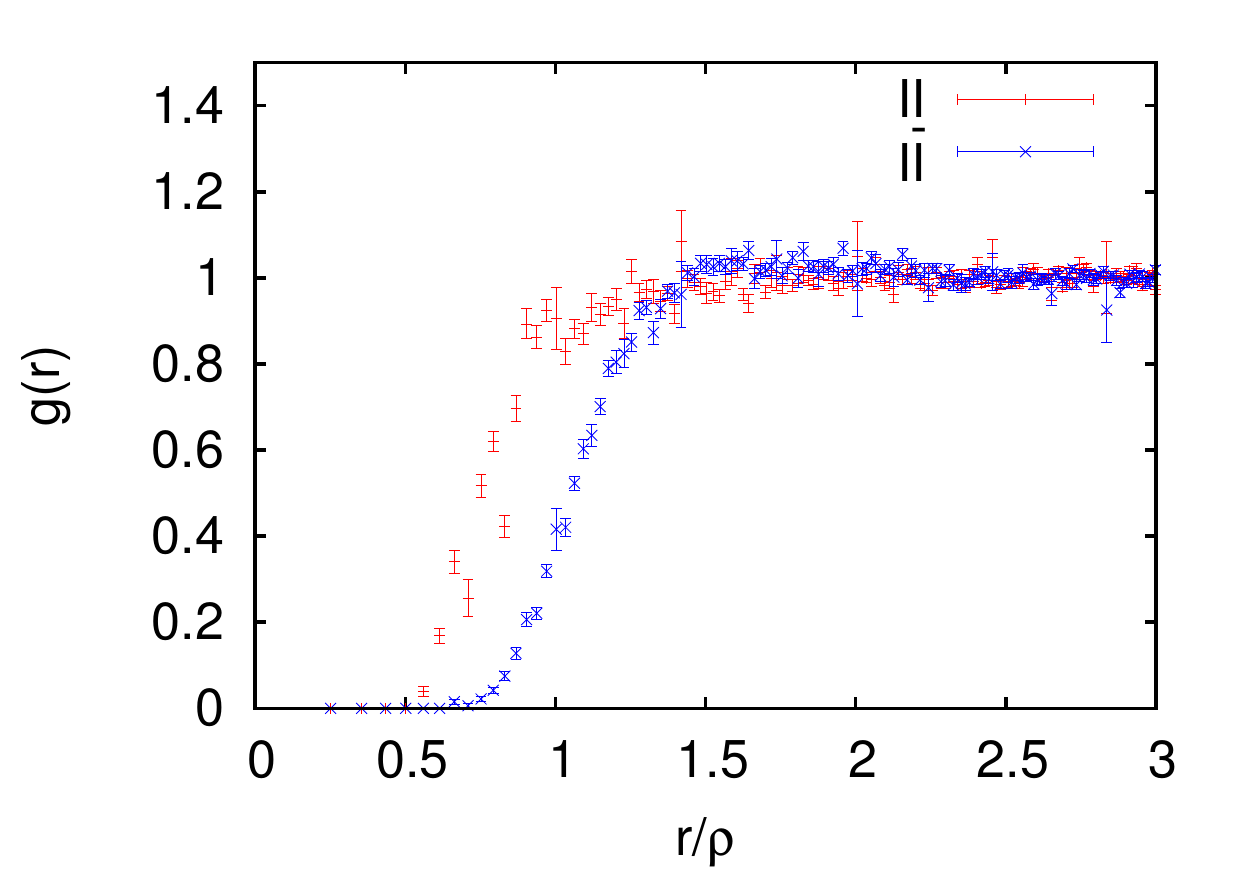} 
\end{tabular} \\
\includegraphics[width=0.5\textwidth]{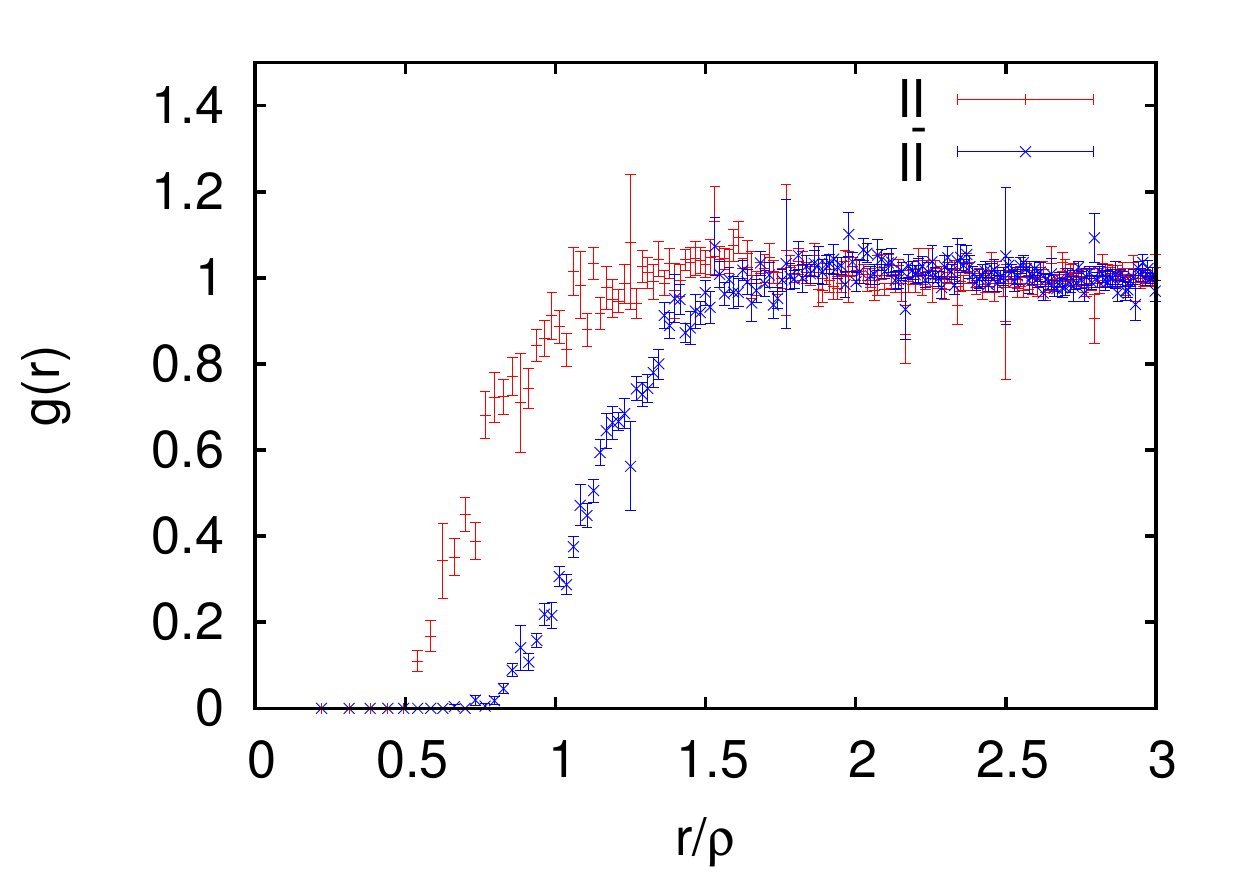}
\end{tabular}
\end{center}
\caption{(Color online) Pair correlation function for $II$ and $AA$ pairs (red) and for $IA$ pairs (blue) for $\beta=4.2$, $V=32^4$ and a Wilson flow time of $\tau=$ 4 (leftmost upper), 8 (rightmost upper) and 15 (lower). The distribution is characteristic of a gas, with an excluded volume that is larger for $IA$ pairs than for $II$ or $AA$.}
\label{fig:gr}
\end{figure}

Furthermore, we have examined the correlations among instantons, both of like ($II$ or $AA$) and unlike charges ($IA$), through the evaluation of the radial distribution function of pairs, $g(r)$. This function is defined as the probability density, evaluated at a distance $r=|{\bf x} - {\bf x_i}|$ away from any given instanton located at the position ${\bf x_i}$, of finding one like- or unlike-charge instanton divided by the density of like- or unlike-charge instantons. Isotropy and translational invariance guarantee that, after averaging over all the possible configurations of the instanton ensemble, the distribution is indeed a radial function irrespective of the position of the instanton $x_i$. At large distances and for the homogeneous distribution of an instanton gas or liquid, correlations among instantons tend to vanish and one would be thereupon left with 
$g(r) \to 1$. This radial distribution $g(r)$ can also be understood as the ratio of the number of instantons found in a spherical shell of infinitesimal width $dr$ and a radius $r$ away from each instanton divided by the shell volume and rescaled by the instanton density. Therefore, for a lattice evaluation, as the lattice volume is proportional to the number of lattice sites, one can in practice estimate the radial distribution of pairs by counting the number of like- or unlike-charge instantons at a given distance away from each instanton, perform then the average over all the instantons of the ensemble, divide next by the product of the total number of avalaible lattice sites at the same distance and the instanton density; and, finally, perform the average over the ensemble of lattice configurations. So proceeding, we have been left with the results displayed in Fig.~\ref{fig:gr} for several Gradient flow times. A first feature to be noticed from the plots is that there is an excluded volume for small distances which grows moderately with the flow time. This might be readily thought to be a consequence of applying the double-counting-preventing prescription that filters out the lumps lying within the core of an accepted instanton. However, the prescription is only applied for like-charge instantons, while the same feature is also clearly manifest for those of unlike charge. Thus, we can conclude that the picture of "{\it hard-core}" instantons is fairly consistent with the results we obtain for the correlations of pairs. Furthermore, the increasing of the excluding volume reflects consistently the growing of the instanton radius which is manifest from Fig.~\ref{fig:size_density_Wilson}. A second remarkable feature is that the distribution of instantons is not completely random: after Gradient flow, it is more likely to find a charge of the same sign than a particle of opposite sign at short distances. The latter is consistently accounted for the fact that the annihilation of opposite charge pairs during the flow increases the probability of finding instantons of the same charge close to each other.  

A short comment about the effect of the filtering for small distances introduced in the localization algorithm is furthermore in order. Had we taken a smaller cut on the instanton minimum inter-distance, a large peak at short distances would have emerged for alike charges in Fig.~\ref{fig:size_density_Wilson}, signaling the presence of very close pairs. The question of whether those {\it instanton clusters} are physical or an artifact of the localization algorithm was discussed long ago~\cite{NEGELE199992,PhysRevD.52.4691}, but up to now there is no definite answer. Thus, our estimates for the instanton density with this algorithm might underestimate systematically the actual density of topological objects because of this. Later on, we will return to this point and justify, on the basis of the comparison of the densities obtained by instanton localization and by Green's functions, that this peak for close pairs ought to be indeed neglected as an artifact of the instanton localization algorithm. However, this implies that one needs, to admit a systematic deviation of around 20-30 \%.

The instanton densities we have measured are of the order of $n \sim 1 {\rm fm}^{-4}$ for rather large flow times, a value that agrees with the one that can be set from the gluon condensate $\VG$, and that has served as reference for decades \cite{RevModPhys.70.323}. Nevertheless, if we try to infer the density at zero Gradient flow time from the results in Fig.\ref{fig:size_density_Wilson}, the density results much larger. Although it is difficult to extrapolate back to $\tau=0$ from our results, it points towards an order of magnitude larger (see next section).

Concerning the instanton size, Garcia-Perez et al. \cite{GARCIAPEREZ1994535} found that when cooling with the Wilson action, individual instantons should shrink under a cooling procedure while with overimproved actions they stabilize or grow with the application of cooling. A similar discussion appears in the nice review by Creutz~\cite{Creutz:2010ec} who also analyzes the effects of these types of action modifications to reflection positivity ~\cite{Luscher:1999un, Creutz:2004ir}. 
Although it is not fully known how the use of different filtering techniques may modify the conclusions reached by the aforementioned authors, the fact of being in a dense liquid instead of being isolated is definitely a dramatic change compared to the setup of the previous studies. Indeed we found that most of the instantons we localized grow with the application of the Gradient flow, while only small instantons shrink (and eventually disappear). The latter appears to indicate that, at a first stage in the evolution with the flow time, a sort of effective interaction between the independent pseudo-instantons (used to describe the classical solution) dominates the process. This interaction takes into account the non-linear effects between the pseudo-particles of the Ansatz which minimize the total action and, at least when the density is large, tends to favor both the annihilation of opposite-charge instantons and their growing when the density drops. We have checked that, if the Gradient flow is driven by the Iwasaki gauge action, the observed evolution does not differ from the one described above.

\section{Momentum running, flow evolution and flavor effects}

As we have discussed in Sec.~\ref{sec:classic}, the running of two- and three-gluon Green's functions can be combined to yield a strong coupling definition, the estimate of which, computed from the gauge fields for a semiclassical multi-instanton ensemble leaves us with a very striking signature for the presence of instantons. In the following, we will use this coupling obtained from the lattice gauge fields to {\it detect} instantons by analyzing its running both at large- and small-momenta, as explained in Ref.~\cite{ATHENODOROU2016354}, for all the set-up's of Tab.~\ref{tab:lattice}. We will then compare the obtained results with the ones stemming from instanton localization for the quenched case at $\beta=4.2$ derived in the previous section, and will also study their evolution with the Gradient flow time. Furthermore, we will take advantage of the other lattice set-ups to analyze the evolution of the instanton ensemble parameters with the number of dynamical flavors. 

\subsection{Momentum running and the instanton density}

\begin{figure}[h!]
\begin{center}
\includegraphics[width=0.7\textwidth]{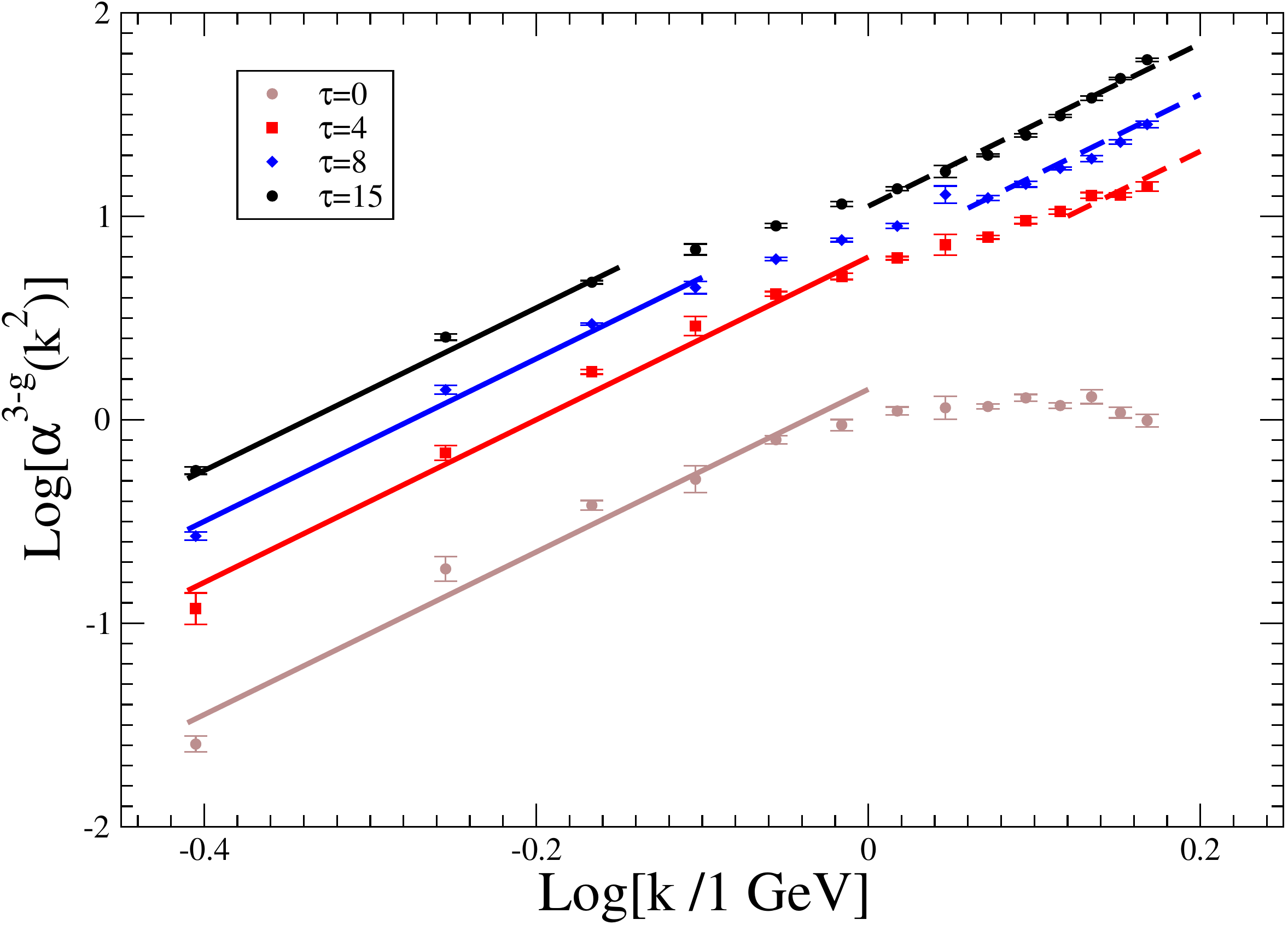} 
\end{center}
\caption{(Color online) The strong coupling defined by \eq{eq:amom}, computed from the lattice gauge fields simulated at $\beta$=4.2 without (brown solid circles) and after applying the Gradient flow with $\tau$=4 (red), 8 (blue) and 15 (black) and depicted by using logarithmic scales. The lines correspond to the multi-instanton ensemble predictions both at low (solid) and large momenta (dashed), according to Eqs.~(\ref{eq:amom},\ref{eq:k4}).}
\label{fig:cool-running}
\end{figure}

Without the application of any filtering technique, the running at large momenta of the strong coupling defined by \eq{eq:amom} is dominated by the perturbative prescription, giving rise to the appearance of asymptotic freedom, but the deep non-perturbative region (typically below $\sim 1{\rm GeV}$) exhibits a behavior that fits well to Eqs.(\ref{eq:amom},\ref{eq:k4}). This finding serves as a confirmation that the low-momenta (large-distance) correlations are dominated by instanton-like objects, and allows to extract the instanton density without the need of any filtering technique. Moreover, after applying the Gradient flow, the UV fluctuations are filtered out and the gluon correlations become again dominated by the semiclassical multi-instanton ensemble and the large-momenta running of the strong coupling is also well accounted by \eq{eq:amom}. This can be clearly seen in Fig.~\ref{fig:cool-running} for the quenched case at $\beta=4.2$, where the lattice estimates appear to be manifestly consistent with \eq{eq:amom}, after applying the Gradient flow, in both the large- and small-momenta domains; thus supporting a multi-instanton ensemble picture for the QCD vacuum at least when describing gluon correlations. It is indeed worthwhile to realize how the suppression of UV fluctuations from the gauge fields turns the well-known large-momenta logarithmic behavior of gluon correlations (brown solid circles in Fig.~\ref{fig:cool-running}, depicting the results from non-flown fields) into the $k^4$-law expressed by \eq{eq:amom}; law which becomes more and more apparent when the Gradient flow time evolves from $\tau$=4 (red circles) to $\tau$=15 (black circles). The same figure had been also presented in \cite{ATHENODOROU2016354}, where its main features were very well understood, in particular how the width for the instanton size distribution shifts the intercept of the logarithmic line up at low momenta, as dicated by \eqref{eq:k4}. 

\begin{table}[h]
\begin{center}
\begin{tabular}{c c c c c c c}
\hline
\hline
$\tau$ & $t/t_0$ & $n\  {\rm (fm^{-4})}$ [IR] &  $n\  {\rm (fm^{-4})}$ [GF] \\
\hline
\hline
 4  & 6.84 & $2.44(34)$ & 3.5(1) \\
 8  & 13.7 & $1.35(18)$ & 1.75(4) \\
 15 & 25.6 & $0.74(8)$ & 0.98(5) \\
\hline
\hline
\end{tabular}
\end{center}
\caption{Comparisons of instanton densities obtained by applying the localization technique (third column) and the Green's function running method (fourth column), after the Gradient flow for three representative times. The errors here quoted incorporate, and are basically dominated by, some systematic uncertainties in the IR case (see text), but only statistical for the GF method. }
\label{tab:comparison}
\end{table}

Now, as the slope of $\alpha_{\rm MOM}(k)$ in $k^4$, according to Eqs.(\ref{eq:amom},\ref{eq:k4}), allows for a direct estimate of the instanton density from the lattice data at large momenta, we consider three representative flow times ($\tau$=4,8 and 15) for the quenched simulation at $\beta$=4.2 and make a systematic comparison of the so obtained densities to the ones resulting from the direct instanton counting using the localization algorithm described in the previous section. The results from both methods lie in the same ballpark, as can be seen in Tab.~\ref{tab:comparison} and Fig.~\ref{fig:extra}. It appears however, that there are systematic deviations which, taking into account the statistical uncertainties, are of the order of 2-3 $\sigma$'s. The deviations can be seen as an underestimation in the counting of instantons by localization of around a 30~\% at $\tau$=4 and of 24~\% at $\tau$=8 and 15. The latter can be consistently understood if one admits that some of the close pairs filtered out by the short-distance criterion correspond to real classical solutions. 
We can estimate that, had we accepted a 20-30 \% of the rejected close pairs, the densities obtained by instanton location would have fairly agreed with the ones resulting from the analysis of the Green's function method. As a fully general classical solution describing such pairs of close instantons is not available, a systematic criterion to disentangle them from single distorted instantons that need to be rejected is not at hand. We have thus preferred to keep the short-distance criterion and assumed a systematic deviation which, anyhow, will not prevent the estimates from being in the same ballpark as those from the Green's function method. 
   
Moreover, either through the analysis of the direct counting of instantons or by the study of the running of  $\alpha_{\rm MOM}(k)$, we reach a coherent image of an instanton density that drops with flow time. It is usually accepted that the results at a small fixed flow time represent the physics of the original gauge field configuration. Nevertheless, the fact that there is such a strong dependence on the flow time requires a deeper understanding. The same phenomenon is found with other filtering techniques such as cooling and we refer the reader to \cite{BOUCAUD2003117} for a simple approach for instanton density evolution through $IA$ annihilation. Let us try here to build a toy model for this phenomenon and use it to extrapolate the instanton density down to zero flow time.

\subsection{Gradient flow evolution of the instanton properties}

If pair annihilation is the only phenomenon responsible for the observed density dropping, the rates at which both $n_I$ and $n_A$ decrease have to be equal. Under the hypothesis that this is a first order process, the time variation of instanton and anti-instanton densities will be given by
\beq
\frac{dn_I}{d\tau} = \frac{dn_A}{d\tau} = - \lambda n_I n_A\ .
\eeq
Furthermore, if $n_I\approx n_A\approx n/2$ (i.e., assuming that the topological charge of a particular gauge configuration is much smaller than the instanton density), we arrive to
\beq
\frac{dn}{d\tau} = -  \frac{\lambda}{2} n^2 \ .
\label{eq:evolution1}
\eeq
The parameter $\lambda$ may depend on the flow time $\tau$ via the instanton size, interparticle distance, etc. thus being not constant. In a first approximation, however, we assume that it is well described by a constant, and integrating out \eq{eq:evolution1} results into
\beq
n(\tau) = \frac{n(0)}{1+\frac{1}{2}\lambda n(0) \tau} = \left( \frac{1}{n(0)} + \frac{\lambda\tau}{2} \right)^{-1}\ .
\label{eq:evolution2}
\eeq

\begin{figure}[h!]
\begin{center}
\includegraphics[width=0.7\textwidth]{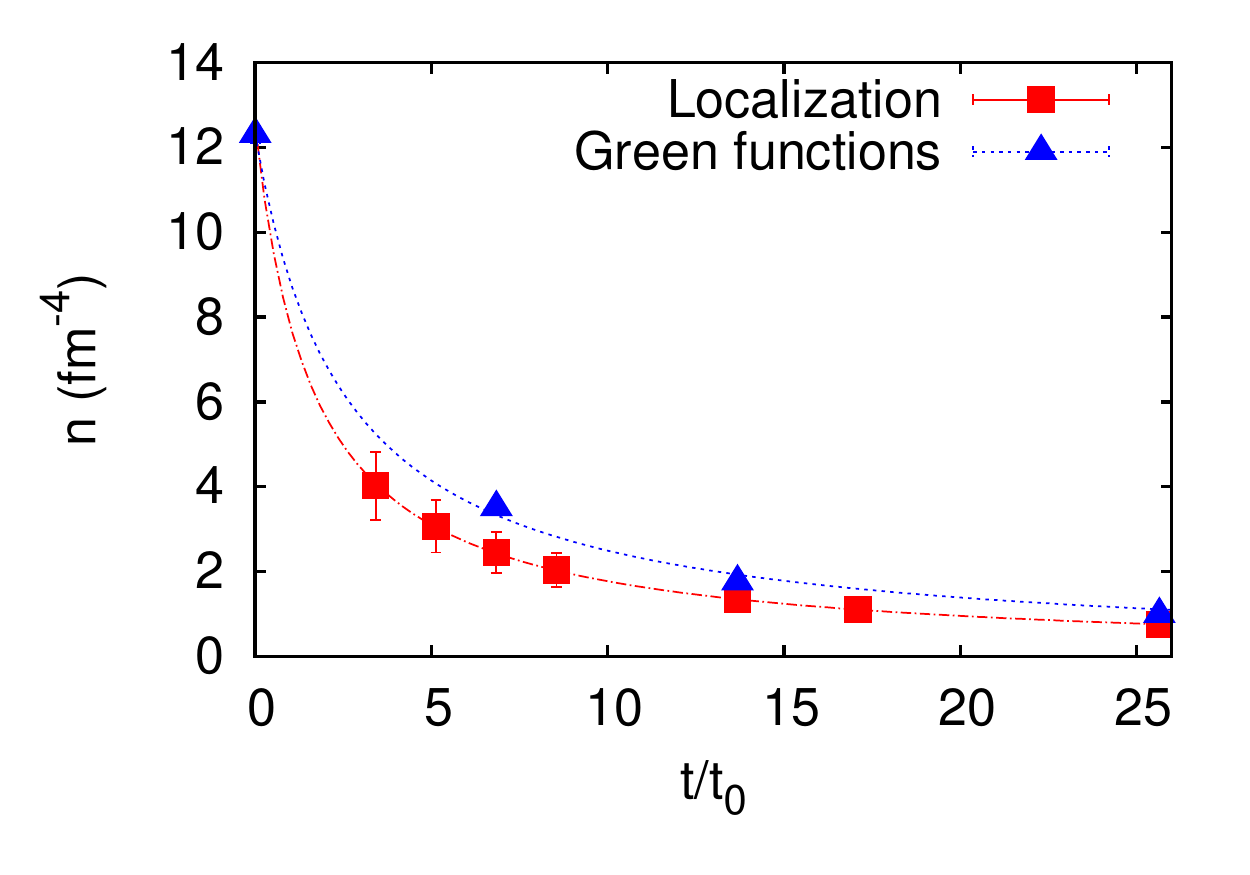}
\end{center}
\caption{(Color online) Comparison of the instanton density evolution with Gradient flow time (in units of $t_0=0.01125 {\rm fm^2}$) from instanton localization (red) and from $\alpha_{\rm MOM}(k)$  (blue). The dashed lines correspond to \eq{eq:evolution2} resulting from the toy model described in the text which fits either localization (red) or Green's functions estimates (blue). The value at zero flow time extrapolated from the instanton localization estimates strikingly coincides with the result from the Green's functions.} 
\label{fig:extra}
\end{figure}

This equation has been used to fit the measured instanton densities at different flow times with the instanton location algorithm for the quenched configurations at $\beta=4.2$, and extrapolate to zero flow time. The result of the fit has been plotted in Fig.~\ref{fig:extra}, thus obtaining an extrapolated density of $12.3(4) {\rm fm}^{-4}$. It is remarkable, despite of the simplicity of the toy model,  how well the measured densities at different flow times agree with \eq{eq:evolution2}. Furthermore, albeit the systematic uncertainty associated to this extrapolation may be rather large and has not been quantified at this level, the so obtained density compares fairly well with the estimate from the Green's function method applied to non-flown gauge fields. The density, in this case, can be obtained once the instanton size distribution width is known and plugged into Eqs.~(\ref{eq:amom},\ref{eq:k4}). This width can be always estimated, after applying the Gradient flow, by the direct scrutiny of the size distribution for the localized instantons (see Fig.~\ref{fig:size_density_Wilson}). However, as discussed in \cite{ATHENODOROU2016354}, it can be evaluated too, though indirectly, from the shift of the low-momentum line with respect to the large-momentum one in the logarithmic plot of Fig.~\ref{fig:cool-running}, according to \eq{eq:k4}, thus providing with a width estimation fully consistently made within the Green's function approach. Owing to its consistency, we have preferred the latter (the estimates from both methods being anyhow in the same ballpark; namely, $\delta\rho^2/\bar\rho^2 $ around 0.02). Furthermore, whilst an extrapolation to zero flow time would be required, in both cases, the values extracted at low flow time remain very stable and we made the simple choice of keeping that at $\tau$=2. Thus, a ratio $\delta\rho^2/\bar\rho^2\approx 0.014$ is estimated and, applied to  Eqs.~(\ref{eq:amom},\ref{eq:k4}), results in $n\approx 12(2) {\rm fm}^{-4}$. As can be also seen in Fig.~\ref{fig:extra}, the four values for the density from Green's functions can be also well described by \eq{eq:evolution2} but with a parameter $\lambda$ slightly diminished.

\subsection{Evolution with the number of flavors}

\begin{figure}[h!]
\begin{center}
\begin{tabular}{c}
\begin{tabular}{cc}
\includegraphics[width=0.52\textwidth]{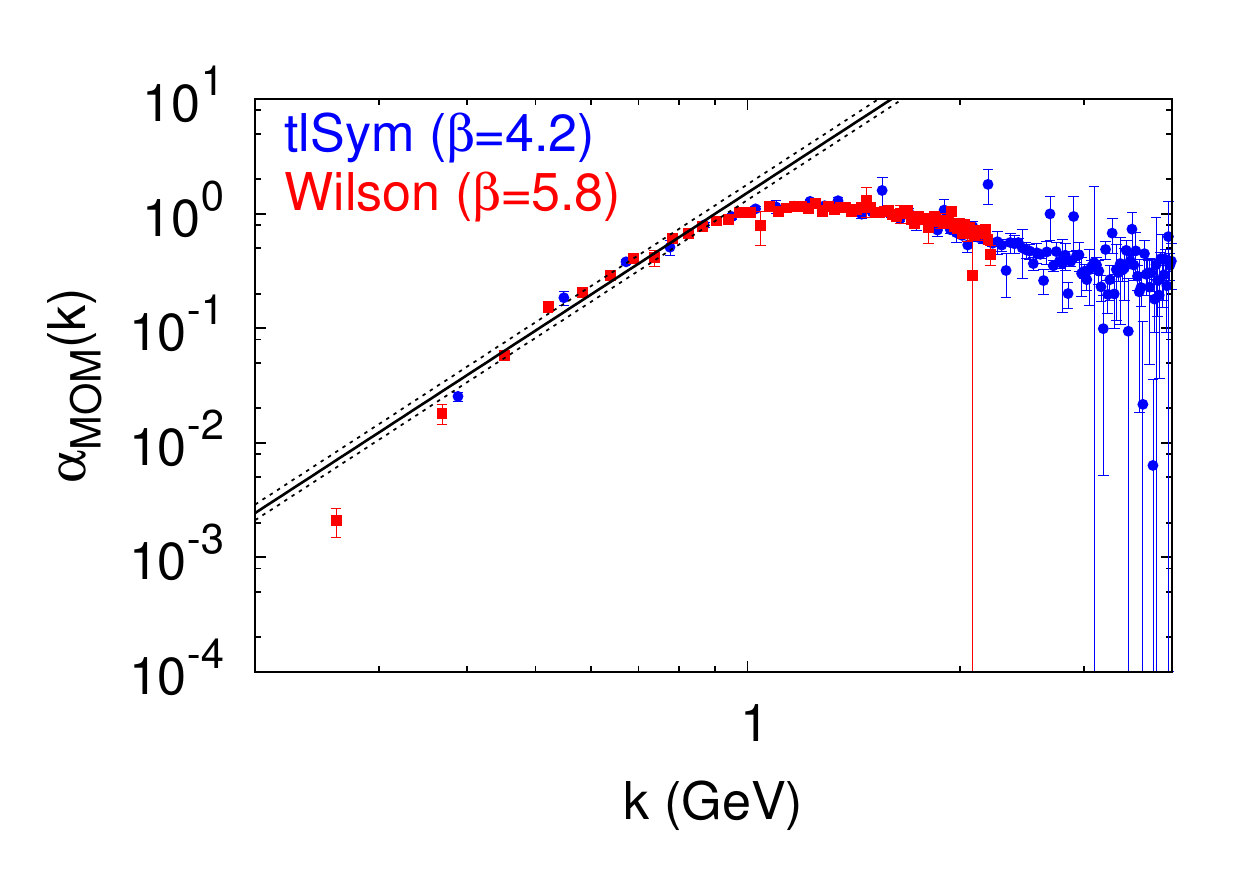} & 
\includegraphics[width=0.52\textwidth]{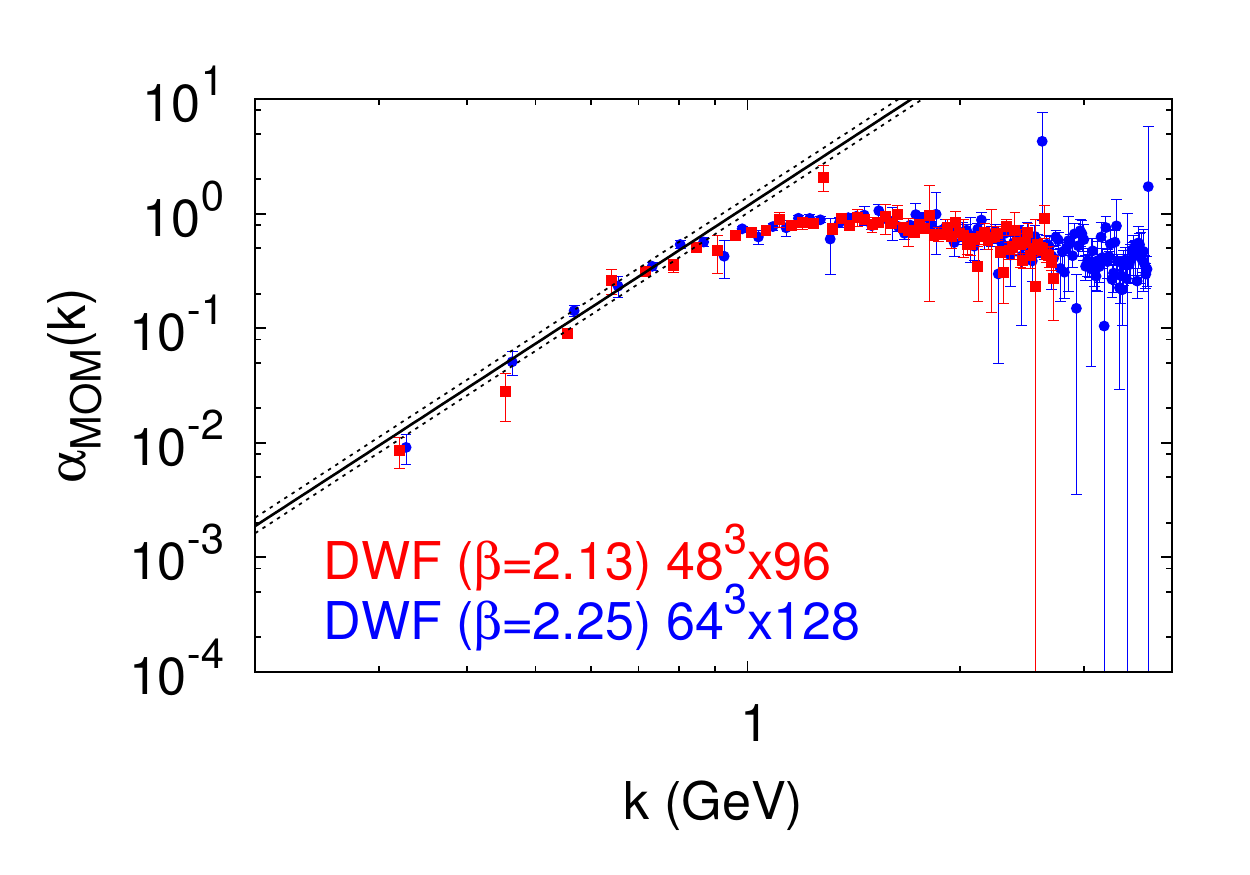} 
\end{tabular} \\
\includegraphics[width=0.52\textwidth]{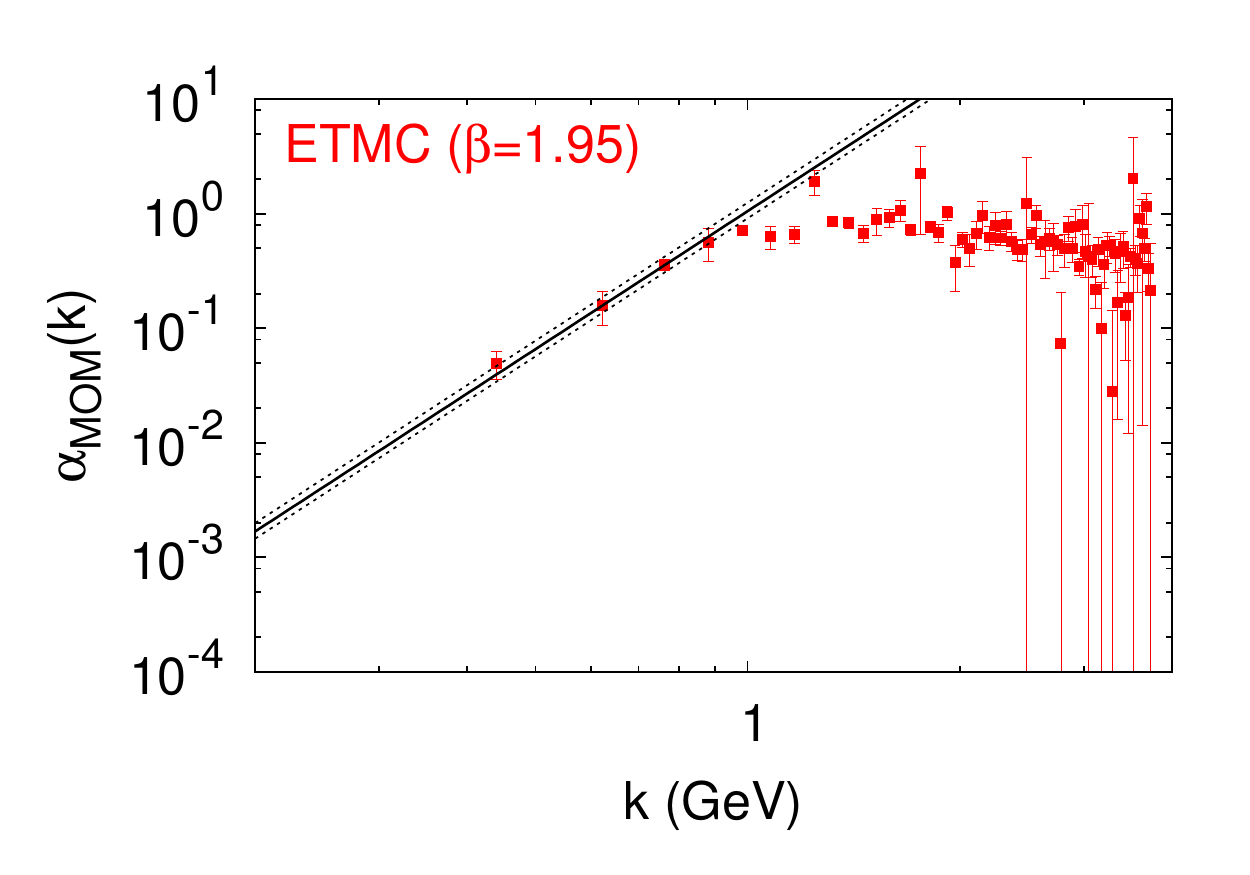} 
\end{tabular}
\end{center}
\caption{(Color online) $\alpha_{\rm MOM}$ vs the momenta obtained from quenched gauge configurations (lefmost upper panel), $N_f=2+1$ Domain-Wall (rightmost upper) and ETMC $N_f=2+1+1$ (lower). The fit to the instanton prediction is represented by the straight line.}
\label{fig:alpha}
\end{figure}

We finally focus on the extensive analysis for the small-momenta behavior of $\alpha_{\rm MOM}(k)$ for the different number of flavors in Tab.~\ref{tab:lattice}. The running of $\alpha_{\rm MOM}(k)$ obtained from the lattice data in Tab.~\ref{tab:lattice}, has been plotted in Fig.~\ref{fig:alpha} for $N_F=0$ (leftmost upper panel), $N_F=2+1$ (rightmost upper) and $N_F=2+1+1$ (lower) without Gradient flow. 
The same instanton picture that, so far, we have firmly grounded on the basis of our deep analysis of quenched lattice simulations ought to work also for the unquenched simulations and, indeed, the running of $\alpha_{\rm MOM}(k)$ between $0.3$ and $0.9$ GeV fits well to the power law given by \eq{eq:k4} (straight line in the plots). When the lattice volume is large enough and there are lattice points at lower momenta (in particular in the quenched $\beta$=5.8 case), this instanton prediction breaks down, presumably due to the neighborhood of a zero crossing for the three gluon vertex ~\cite{Aguilar:2013vaa,Tissier:2011ey,Pelaez:2013cpa,Blum:2014gna,Eichmann:2014xya,Cyrol:2016tym,ATHENODOROU2016444,Duarte:2016ieu,PhysRevD.95.114503}. From the point of view of the ILM, the failure for small momenta is expected because for very large distances (typically larger than the instanton size) the model of uncorrelated instantons is not expected to work; in any case the purely quantum nature of the zero-crossing discussed in Refs.~\cite{ATHENODOROU2016444,PhysRevD.95.114503} seems to be in contradiction with an instantonic explanation. 

Now, as \eq{eq:amom} tells, the IR running of $\alpha_{\rm MOM}(k)$ is basically determined by the instanton density and, thus, a direct comparison of the results of the low-momenta fits shown in Fig.~\ref{fig:alpha} makes possible the scrutiny of the evolution of this density with the number of flavors. Indeed, it suggests that the instanton density increases monotonously with the number of dynamical flavors. If we furthermore accept that the ratio $\delta\rho^2/\bar\rho^2$ is the same for $N_f=0$, $N_f=2+1$ and $N_f=2+1+1$, we obtain that the instanton density increases by a factor $1.3(1)$ from $N_f=0$ to $N_f=2+1$ and by a factor $1.5(1)$ from $N_f=0$ to $N_f=2+1+1$. These two ratios are nearly compatible within the errors and this suggests a very mild effect of the charm quark on the instanton density, certainly owing to its sizeable mass threshold. The role of the charm quark and the dependence on the quark masses has not been studied in detail yet but 
the fact that the instanton density grows with the presence of dynamical flavors has been previously reported, for instance, in Ref.~\cite{PhysRevD.78.054506}.  In the previous reference it was interpreted as follows: it is well known that an isolated instanton produces a zero mode of the fermion determinant~\cite{HART2001280,PhysRevD.78.054506} and, as a consequence of this and the action of a given gauge field being weighted by $[\mathrm{det} (\slash \hspace*{-0.25cm} D +m)]^{N_F}$, the existence of zero modes of the Dirac operator would be highly improbable in the chiral limit.  Becoming more improbable the larger is the number of light fermion flavors. Thereupon, a large density of instantons can be understood as favoring the instanton superposition and thus suppressing the effect of isolated instantons. Or, in other words, it can be thought to enhance non-linear effects and pseudo-particle interactions destroying the zero-modes associated to the single instanton solution. However, one needs to be careful with the above argument (that for the massless theory, one can ignore
any instanton effects since those configurations don't contribute to the partition function) since while it is correct that instantons do drop out of the partition function itself they do survive in physical correlation functions. We refer the reader to ~\cite{Creutz:2005rd} for a detailed analysis of the topic and its relation to the 't Hooft vertex~\cite{tHooft:1976rip}.  

On the other hand, even admitting that the above qualitative argument might work, there is another source of explanation, at least partially, for this increasing with the number of dynamical flavors: any direct comparison of the instantons densities obtained from simulations with different number of flavors gets over the systematic deviations from the physical scale setting on phenomelogical estimates or in gauge quantities as the gluon correlation functions~\cite{Boucaud:2017ksi,Duarte:2017wte}. In setting the scale by imposing that a lattice estimate of a given quantity takes its physical value, the latter always incorporates effects from those dynamical quarks being active in the real world for the empirical determination of this quantity. In particular, the impact of this effect on a quenched theory is certainly non-negligible, but still for $N_f$=2 simulations, there has been a recent work~\cite{Binosi:2016xxu} finding a reduction of around a 5~\% in the lattice spacing as a systematic deviation from the physical scale setting. A reduction of the lattice spacing ought also to be expected for the quenched lattice theory, as the lattice estimates for $\Lambda_{\overline{\rm MS}}$ appear to be lower at $N_f$=0~\cite{Boucaud:2008gn} than at $N_f$=2 or $N_f$=2+1+1~\cite{Aoki:2016frl,Blossier:2012ef}, while the matching at the quark-mass thresholds\footnote{A matching of theories with $N_f$ and $N_f$+1 dynamical flavors at the quark-mass threshold (or a similar mechanism~\cite{Binosi:2016xxu}) is, precisely, what is required for a meaningful comparison of quantities derived from those theories.} of theories with different number of dynamical flavors suggest the opposite trend~\cite{Agashe:2014kda}.  Though it is very hard to quantify how this reduction amounts, one can readily conclude that a diminishing of a 10~\% for the lattice spacing in the quenched theory would enhance by around a 40~\% the instanton density and would thus make it lie on the same ballpark as $N_f$=2+1 and $N_f$=2+1+1. There might appear to exist also an impact from the charm quark in the lattice scale setting for the comparison of $N_f$=2+1 and $N_f$=2+1+1 results, however the scale setting involving quantities in the pion sector, such an impact should be nearly negligible~\cite{Binosi:2016xxu}.  

Finally, one must also keep in mind that the estimates for the densities we compare have been obtained by invoking Eqs.~\eqref{eq:amom} and \eqref{eq:k4}, the last of which also needs the instanton size width to be plugged into. We have assumed, for the comparison to be made, that $\delta \rho^2/\bar{\rho}^2$ is the same for any number of flavors. Whilst the latter seems to result, at low Gradient flow time, from the analysis of Ref.~\cite{ATHENODOROU2016354} and despite the fact that the width is only playing at the level of a subleading correction, any small deviation would become magnified by the factor $48$ in \eqref{eq:k4} and so produce a sizeable impact on the comparison. However, as above stated, the picture we obtain for the impact of the dynamical flavors on the instanton density is consistent with the results of Ref.~\cite{PhysRevD.78.054506}, obtained with a different technique for the instanton detection.

\section{Conclusions}
\label{sec:conclusions}

We have presented an analysis of the instanton contributions to the QCD vacuum using two complementary approaches. The first one is based on the fact that an independent pseudo-particle approach (i.e. an instanton liquid model without correlations among instantons) predicts a power law $k^4$ for the combination of gluon Green's functions used to define the coupling in the symmetric ${\rm MOM}$ scheme, $\alpha_{\rm MOM}(k)$. Moreover, the coefficient of $k^4$ can be used to infer the instanton density. The second approach uses the Gradient flow technique to remove short range fluctuations in the gauge field configurations and instanton-like structures are revealed in the topological charge density. We have proposed a method for identifying instantons among the extrema of the topological charge density and applied it to some quenched configurations. 

Although GF is thought to modify minimally the topological structure of the gauge field configuration, it does not prevent the disappearance of instantons, neither of small instantons whose size is comparable to the lattice spacing, nor through instanton/anti-instanton annihilation. Indeed, we have observed that both methods (IR running of $\alpha_{\rm MOM}(k)$ and instanton localization) show a strong decrease in the instanton density with the flow time. Whilst the densities obtained from both methods differ by a $\sim 25-30~\%$, they can be seen to lie in the same ballpark  and, on top of this, the fact that both evolve in the same manner with flow time strongly supports the instanton dominance after removing the short range fluctuations. The existence of noticeable discrepancies between the instanton densities obtained from both methods can be well interpreted as due to the systematic uncertainties related to the difficulties for identifying close instanton pairs (or clusters) with our localization algorithm owing to strong deviations locally around the instanton centers from the BPST profile used. 

A simple model for instanton annihilation reproduces,  qualitatively and quantitatively, the evolution of the instanton with GF, and allows to extrapolate the instanton density to zero flow time. The density obtained, $\sim 12 {\rm fm}^{-4}$, is much larger than the traditionally quoted value of $1  {\rm fm}^{-4}$, although some modern estimates~\cite{PhysRevD.88.034501} point towards larger densities. For zero flow time, i.e. without the application of the Gradient flow, we cannot localize instantons using our direct method, but can still fit the instanton density using the $k^4$ power law. We obtained that the slope of $\alpha_{\rm MOM}(k)$ supports the picture of a dense instanton liquid, with a density of $n\approx 12.3(2) {\rm fm}^{-4}$. This density is fully compatible with the extrapolation of the densities obtained from the localization method using the annihilation model.

After performing the comparison of both methods in the quenched case, we obtained the instanton density for unquenched ($N_F=2+1$ and $N_F=2+1+1$) lattice gauge field configurations without GF. From the fit to the power-law, we obtained larger instanton densities for the unquenched case, which although already reported \cite{PhysRevD.78.054506} is a somehow controversial finding. The naive argument associates, via the Atiyah-Singer index theorem, instantons to zero modes of the Dirac operator and, thus, the presence of light dynamical quarks in the unquenched simulation should suppress instantons. Indeed our numerical findings with both methods employed seem to indicate the opposite behavior, namely the density of instantons grows so that they get distorted and can not any longer be associated to zero modes of Dirac operator. Two considerations can be made concerning this enhancement of the instanton density with the inclusion of dynamical quarks: first only the light quarks should have an effect and, therefore, the inclusion of the charm quark when passing from $N_F=2+1$ to $N_F=2+1+1$ is thought to have a negligible impact over the instanton density. We found ratios to the quenched case of $1.3(1)$ for $N_F=2+1$ and $1.5(1)$ for $N_F=2+1+1$, nearly compatible within errors. The second consideration discussed in the text is the impact of the light quarks in the quenched lattice spacing setting. When using any observable to fix the lattice spacing, it incorporates the effect of any quark being active at the scale used for the determination of the physical observable.
Therefore, in the comparison of quenched to unquenched results, it is conceivable that a systematic deviation of the lattice spacing is present. For our estimates of the instanton density, a $\sim 10\%$ systematic deviation for the quenched lattice spacing would account for the differences between $N_F=0$ and $N_F=2+1$ and $2+1+1$. 

In summary, we feel that the results presented here draw a clear image of instanton dominance after application of gradient flow, when there is no trace of $\Lambda_{\rm QCD}$, and the running of gluon Green's functions is dominated by the semiclasical content of the gauge fields. Perhaps more interestingly, the IR behavior of gluon Green's functions without gradient flow seems to be well described by an instanton liquid model with a rather large density. Why instantons may serve for describing some properties of low energy QCD and not others, or to what extent instantons are relevant for the QCD phenomenology seem to be still open problems.

\section*{Acknowledgements} 

We thank the support of Spanish MINECO FPA2014-53631-C2-2-P  and FPA2017-86380-P research projects. 
SZ acknowledges support 
by the National Science Foundation (USA) under grant
PHY-1516509, by the Jefferson Science Associates,
 LLC under  U.S. DOE Contract \#DE-AC05-06OR23177 and by the DFG Collaborative Research Centre SFB 1225 (ISOQUANT).
We acknowledge fruitful discussions with M. Creutz, M. Garcia-Perez, K. Orginos, E. Shuryak, I. O. Stamatescu, T. Sulejmanpasic and J. Verbaarschot. We are also extremely thankful to the ETM and RBC/UKQCD collaborations for sharing their gauge configurations with us.
Numerical computations have used resources of CINES, GENCI-IDRIS   under the
allocation 52271, and of the IN2P3 computing facility in France as well as of L-CSC in Germany.



\providecommand{\href}[2]{#2}\begingroup\raggedright\endgroup

\end{document}